\documentclass{aa}
\usepackage[varg]{txfonts}

\usepackage[english]{babel}
\usepackage[utf8]{inputenc}
\usepackage{natbib}
\usepackage{xspace}
\usepackage{siunitx} 
\usepackage{color}
\usepackage{graphicx}
\usepackage{esvect}
\usepackage{makecell}
\usepackage{tabularx} 
\usepackage{multirow}
\usepackage{multicol}
\usepackage{pdflscape}
\usepackage{subcaption}
\usepackage{setspace}
\usepackage{xcolor}
\usepackage{pdfpages}

\newcommand*\solarmass{\si{\solarmass}\xspace}
\DeclareSIUnit\solarmass{\ensuremath{\textnormal{M}_{\odot}}}
\newcommand*\solarmetal{\si{\solarmetal}\xspace}
\DeclareSIUnit\solarmetal{\ensuremath{\textnormal{Z}_{\odot}}}
\newcommand*\parsec{\si{\parsec}\xspace}
\DeclareSIUnit\parsec{pc}
\newcommand*\erg{\si{\erg}\xspace}
\DeclareSIUnit\erg{erg}
\newcommand*\yr{\si{\yr}\xspace}
\DeclareSIUnit\yr{yr}
\newcommand*\rfivehundred{\si{\rfivehundred}\xspace}
\DeclareSIUnit\rfivehundred{\ensuremath{\textnormal{r}_{500}}}
\newcommand*\pixel{\si{\pixel}\xspace}
\DeclareSIUnit\pixel{pixel}

\newcommand*\ie{i.e.\ } 
\newcommand*\eg{e.g.\ } 

\newcommand*\erosita{eROSITA\xspace}
\newcommand*\hiflugcs{HIFLUGCS\xspace}
\newcommand*\rass{RASS\xspace}
\newcommand*\rosat{ROSAT\xspace}
\newcommand*\chandra{\texttt{Chandra}\xspace}

\newcommand*\numcolor{\textcolor{black}}

\def\url#1{\expandafter\string\csname #1\endcsname}

\graphicspath{
{./anc/},
{./anc/scaling_relations/},
{./anc/emprofiles/},
{./anc/images/},
{./anc/surfbri/},
{./anc/ellipticity/},
{./anc/residuals/},
{./anc/mass_comparisson/}
}

\begin{document}
\renewcommand{\arraystretch}{1.5}

\title{Towards a characterization of X-ray galaxy clusters for cosmology} 


\author{
Florian K{\"a}fer\inst{1}
\and Alexis Finoguenov\inst{1,2}
\and Dominique Eckert\inst{1,3}
\and Jeremy S. Sanders\inst{1}
\and Thomas H. Reiprich\inst{4}
\and Kirpal Nandra\inst{1}
} 

\offprints{F. K{\"a}fer, \email{fkaefer@mpe.mpg.de}}

\institute{
Max-Planck-Institut f{\"u}r extraterrestrische Physik, Giessenbachstra{\ss{}}e, 85748 Garching, Germany
\and
Department of Physics, University of Helsinki, PO Box 64, 00014, Helsinki, Finland
\and
Department of Astronomy, University of Geneva, Ch. d'Ecogia 16, 1290 Versoix, Switzerland
\and
Argelander-Institut f{\"u}r Astronomie, Universit{\"a}t Bonn, Auf dem H{\"u}gel 71, 53121 Bonn, Germany
}

\date{Received / Accepted 21st of June 2019}

\abstract
{
In the framework of the hierarchical model the intra-cluster medium  properties of galaxy clusters are tightly linked to structure formation, which makes X-ray surveys well suited for cosmological studies. To constrain cosmological parameters accurately by use of galaxy clusters in current and future X-ray surveys, a better understanding of selection effects related to the detection method of clusters is needed.
}
{
We aim at a better understanding of the morphology of galaxy clusters to include corrections between the different core types and covariances with X-ray luminosities in selection functions. In particular we stress the morphological deviations between a newly described surface brightness profile characterization and a commonly used single $\beta$-model.
}
{
We investigate a novel approach to describe surface brightness profiles, where the excess cool-core emission in the centres of the galaxy clusters is modelled using wavelet decomposition. Morphological parameters and the residuals are compared to classical single $\beta$-models, fitted to the overall surface brightness profiles.
}
{
Using single $\beta$-models to describe the ensemble of overall surface brightness profiles leads on average to a non-zero bias ($0.032 \pm 0.003$) in the outer part of the clusters, \ie a $\sim 3\%$ systematic difference in the surface brightness at large radii. In addition $\beta$-models show a general trend towards underestimating the flux in the outskirts for smaller core radii. Fixing the $\beta$ parameter to $2/3$ doubles the bias and increases the residuals from a single $\beta$-model up to more than $40\%$. Modelling the core region in the fitting procedure reduces the impact of these two effects significantly. In addition, we find a positive scaling between shape parameters and temperature, as well as a negative correlation ($\sim-0.4$) between extent and luminosity.
}
{
We demonstrate the caveats in modelling galaxy clusters with single $\beta$-models and recommend the usage of them with caution, especially when not taking the systematics into account that arise using them. Our non-parametric analysis of the self-similar scaled emission measure profiles indicates no systematic core-type differences of median profiles in the galaxy clusters outskirts.
} 

\keywords{X-rays: galaxies: clusters -- general -- cosmology: observations}

\maketitle

%
\section{Introduction}
\label{sec:introduction}

Clusters of galaxies are formed from the collapse of initial density fluctuations in the early Universe and grow hierarchically to the densest regions of the large-scale structure. This makes them the most massive ($M_{\text{tot}}\sim$\SIrange[range-units = single, range-phrase=--]{E14}{E15}{\solarmass}) gravitationally bound structures in our universe and their virialisation timescales are less than the Hubble time. The gas between the galaxies, the intra-cluster medium (ICM), has been heated  to temperatures\footnotemark{} of several \SI{E7}{\K} by gravitational collapse. The primary emission mechanism of this hot, fully-ionised thermal plasma is thermal bremsstrahlung and line emission of heavy elements, \eg iron. The majority ($\sim$85 \%) of the baryonic component is in the form of the hot ICM. Therefore the most massive visible component can be traced by X-ray emission, which makes X-ray astronomy a great and important tool to study galaxy clusters.
However, flux-limited  galaxy cluster samples compiled from X-ray surveys suffer from selection effects like Malmquist bias, \ie the preferential detection of intrinsically brighter sources (a more detailed discussion of different selection effect biases is compiled in \eg \citealt{2010A&A...513A..37H,2013SSRv..177..247G}). Another form of selection effect arises from the different core-types of galaxy clusters. In the central regions of galaxy clusters, gas is able to cool more efficiently compared to the outskirts. Several diagnostics were proposed to identify and categorize galaxy clusters according to their different core-types, \eg a central temperature decrease \citep{2006MNRAS.372.1496S}, mass-deposition rates \citep{2007A&A...466..805C}, cuspiness \citep{2007hvcg.conf...48V} or surface brightness concentration \citep{2008A&A...483...35S}. Galaxy clusters exhibiting cool-cores show centrally-peaked surface brightness profiles, whereas non-cool-core clusters have flat profiles. In surveys differently shaped profiles are detected with different efficiencies. Even for the same brightness, cool-core clusters may be more easily detected, since their surface brightness profiles are more peaked and thus the central emission sticks out more above the background. The preferential detection of cool-core objects close to the detection threshold of flux-limited samples leads to the so-called cool-core bias \citep[e.g.][]{2011A&A...526A..79E, 2017MNRAS.468.1917R}.
It is crucial to take such selection effects into account in cosmological studies to obtain unbiased results. One possibility to quantify these biases is running the source detection chain on well-defined simulations. The quantification of the completeness, \ie the fraction of detected clusters as function of mass and redshift, requires an accurate galaxy cluster model as input for such simulations.
Outside the core regions, scaled radial profiles (\eg temperature, pressure or entropy profiles) of galaxy clusters show a so-called "self-similar" behaviour \citep[e.g.][]{2007A&A...467..437Z,2018arXiv180500042G}. It is believed that this is the result of a similar formation process of galaxy clusters, namely that tiny density perturbations in the early universe are amplified by gravitational instabilities and grow hierarchically, yielding the large scale structure observed today. Galaxy clusters are then believed to correspond to the densest regions of the large scale structure. This formation history motivated the theoretical consideration of the self-similar model \citep[e.g.][]{1986MNRAS.222..323K}, where all galaxy clusters share the same average density and evolve with redshift and mass according to prescriptions given by spherical gravitational collapse. Therefore galaxy cluster observables such as X-ray luminosity, spectral temperature or gas mass are correlated to the total cluster mass. Assuming gravity is the dominant process, the self-similar model predicts simple power-law relations between those cluster observables and the total mass, so-called scaling relations \citep[e.g.][]{2007ApJ...668..772M,2009A&A...498..361P,2010MNRAS.406.1773M,2016MNRAS.463.3582M,2012MNRAS.421.1583M}.
In this work, we aim towards a proper characterization of galaxy cluster shapes using different surface brightness parameterisations. We investigate scaling relations between surface brightness parameters and temperature by use of the HIghest X-ray FLUx Galaxy Cluster Sample (\hiflugcs), a statistically complete, X-ray-selected and X-ray flux-limited sample of $64$ galaxy clusters compiled from the ROSAT all-sky survey (\rass, \citealt{1999A&A...349..389V}). In addition we study the covariances between shape and other galaxy cluster parameters. The impact of the different core-types on the obtained scaling relations and covariances are quantified.
The goal of this paper is to improve our understanding of galaxy cluster shapes. This serves as a basis for simulations quantifying selection effects, inter alia for the future X-ray all-sky survey performed by the extended Roentgen Survey with an Imaging Telescope Array (\erosita, \citealt{2012arXiv1209.3114M,2018SPIE10699E..5HP}). In addition, the obtained covariance matrices can be implemented in current cosmological studies using \eg the COnstrain Dark Energy with X-ray galaxy clusters (CODEX) sample.
Throughout this paper a flat $\Lambda$CDM cosmology is assumed. The matter density, vacuum energy density and Hubble constant are assumed to be $\Omega_{\textnormal{m}} = \num{0.3}$, $\Omega_{\Lambda} = \num{0.7}$ and $H_{0} =$ \SI{70}{\kilo\metre\per\second\per\mega\parsec} with $h_{70} \coloneqq H_{0} / \SI{70}{\kilo\metre\per\second\per\mega\parsec} = \num{1}$, respectively. 'ln' refers to the natural logarithm and 'log' is the logarithm to base 10. All errors are $1\sigma$ unless otherwise stated.

\footnotetext{X-ray gas temperatures are often expressed in kiloelectronvolt (\SI{}{\keV}). Using the Boltzmann constant $k_{\textnormal{B}}$, $\SI{1}{\keV} / k_{\textnormal{B}} \approx \SI[exponent-product = \cdot]{1.16E7}{\K}$.}

%
\section{Data}
\label{sec:data}

This section describes the underlying galaxy cluster sample used for this study, as well as a description of the data analysis.

\subsection{The sample}

The HIghest X-ray FLUx Galaxy Cluster Sample \citep[\hiflugcs,][]{2002ApJ...567..716R,2010A&A...513A..37H} comprises $64$ galaxy clusters, constructed from highly complete cluster catalogs based on the \rosat all-sky survey (\rass). The final flux limit of $f_{\text{X}}\text{(\SIrange[range-units = single, range-phrase=--]{0.1}{2.4}{\keV})} = \text{\SI[exponent-product = \cdot]{20E-12}{\erg\per\second\per\square\centi\metre}}$ defines the X-ray-selected and X-ray flux-limited sample of the brightest galaxy clusters away from the Galactic plane. Although statically complete, \hiflugcs is not necessarily representative or unbiased with respect to the cluster morphology \citep{2010A&A...513A..37H,2011A&A...532A.133M}. \citet{2011A&A...526A..79E} calculated a significant bias in the selection of X-ray clusters of about 29 percent in favour of centrally peaked, cool-core objects compared to non-cool-core (NCC) clusters. We minimize these kind of selection effects by restricting our study to objects above a temperature of \SI{3}{\keV}. This temperature threshold excludes the low mass galaxy groups, which are closer to the \hiflugcs flux threshold and therefore have a high cool-core fraction as discussed in Sect. \ref{sec:introduction}. Originally, the cluster RX J1504.1-0248 was not included in \hiflugcs. This object is a strong cool-core (SCC) cluster that appears only marginally extended in the \rass, \ie its extent is comparable to the \rosat survey point-spread function (PSF). To avoid biasing our results because of the small extent of this system compared to the \rosat PSF, this cluster is excluded in our full analysis. In total, we consider \num{49} galaxy clusters above our selected temperature threshold of \SI{3}{\keV}. 

\subsection{Data analysis}

Pointed \rosat observations were used whenever available. The pointed data is reduced with the ROSAT Extended Source Analysis Software package \citep{1994ApJ...424..714S} as described in \citet{2012A&A...541A..57E}. Otherwise \rass data from the public archive are used. There are \num{4} objects without pointed observations above our representative temperature threshold of \SI{3}{\keV} and excluding those from the analysis does not change our results significantly. Therefore, we do not expect our results to be affected by the use of heterogeneous data. All images for count rate measurements are restricted to the \rosat hard energy band (channels \numrange[range-units = single, range-phrase=--]{42}{201} $\approx$ \SIrange[range-units = single, range-phrase=--]{0.4}{2.0}{\keV}) due to higher background levels in the soft band. Luminosities are taken from \citealt{2002ApJ...567..716R}, \ie are based on \rosat data and are not corrected for cooling flows. The central temperature drop of the ICM in cool-core clusters biases the estimation of the cluster virial temperature, \ie the temperature of the hot gas which is in hydrostatic equilibrium with the potential well of the cluster. This bias is a source of scatter in scaling relations related to temperature and can be minimized by excluding the central region for the temperature fitting. Since we are interested in how the galaxy cluster shape parameters scale with temperature, we adapt core-excised \hiflugcs temperatures measured by \citealt{2010A&A...513A..37H} using \chandra's Advanced CCD Imaging Spectrometer (ACIS) data. We re-scale all temperatures greater than \SI{2}{\keV} due to the \chandra calibration package updates according to the \citet{2011A&A...532A.133M} best-fit relation 
\begin{equation}
T_{4.1.1} = 0.875 \cdot T_{3.2.1} + 0.251,
\label{equ:tcalibration}
\end{equation}
which links temperature measurements (in \SI{}{\keV}) between the
Calibration Database (CALDB) 3.2.1 and CALDB 4.1.1.
The reader is referred to the aforementioned papers in this subsection for a detailed description of individual data analysis steps.

\subsection{Masses}

We adapt $M_{500}$ values of the "Union catalogue" \citep{2016A&A...594A..27P} calculated by Planck Sunyaev-Zel'dovich (SZ) observations, which contains detections with a minimum signal-to-noise of \num{4.5}. We note that SZ mass uncertainties are small due to being purely statistical and are not propagated when rescaling radii. The advantage of using masses calculated by the use of the $Y_{\textnormal{SZ}}-M$ relation in this study is that they are statistically not covariant with X-ray parameters and less affected by galaxy cluster core states \citep{2015ApJ...802...34L}. Above our selected temperature threshold of \SI{3}{\keV} there are \numcolor{4} galaxy clusters without counterpart in the Planck catalogue (Hydra-A, A1060, ZwCl1215, A2052). These clusters are rejected for studies that require characteristic masses or radii. Assuming spherical symmetry, the galaxy cluster masses can be transformed into $r_{500}$ values according to
\begin{equation}
r_{500} = \left( \frac{3 M_{500}} {4 \pi \cdot 500 \rho_{\textnormal{crit},z}} \right)^{1/3}.
\label{equ:r500}
\end{equation}
%

%
\section{Analysis}
\label{sec:analysis}

\subsection{Surface brightness profiles}

Using \citeauthor{1962AJ.....67..471K}{\color{blue}'s} (\citeyear{1962AJ.....67..471K}) analytical approximation of an isothermal sphere, measured X-ray surface brightness profiles of galaxy clusters are well described\footnotemark{} by a $\beta$-model \citep{1976A&A....49..137C}
\begin{equation}
s_{\textnormal{X}}(R) = \sum_{j=1}^{N} s_{0,j} \left[ 1 + \left( \frac{ R } { r_{\textnormal{c},j} } \right)^{2} \right]^{-3 \beta_{j} + \num{0.5}}.
\label{equ:betamodel}
\end{equation}
For each component $j$, $s_{0,j}$ is the central surface brightness, \ie at projected radial distance $R=0$, $r_{\textnormal{c},j}$ is the core radius of the gas distribution and the slope $\beta_{j}$ is motivated by the ratio of the specific energy in galaxies to the specific energy in the hot gas. For galaxy clusters exhibiting a central excess emission due to the presence of cool cores, a double ($N=2$) $\beta$-model can improve the agreement between model and data as one component accounts for the central excess emission while the other accounts for the overall cluster emission. However, the two components are highly degenerate and except for very nearby galaxy clusters, the \rosat point-spread function is insufficient to resolve the core regions since the apparent size of the objects is smaller. Therefore, a single ($N=1$) beta model is used to describe the galaxy cluster emission and the central excess emission is included in the background model.
Simulations \citep{1995MNRAS.275..720N,1996MNRAS.283..431B} indicate that the measured $\beta$ values are biased systematically low if the range of radii used for fitting is less than the virial radius of the cluster. The advantage of using \rosat PSPC data to determine the surface brightness profiles is in the large field-of-view and the low background, allowing to trace the galaxy cluster emission to relatively large radii.
\footnotetext{Note that the assumption of a single $\beta$-models is that the hot gas and the galaxies are in hydrostatic equilibrium and isothermal.}

\subsubsection{Wavelet decomposition}

We use a wavelet decomposition technique as described in \citet{1998ApJ...502..558V}. The technique is implemented as \texttt{wvdecomp} task of the publicly available ZHTOOLS\footnote{Please contact A. Vikhlinin for the latest version of ZHTOOLS (avikhlinin@cfa.harvard.edu).} package. The basic idea is to convolve the input image with a kernel which allows isolation of structures of given angular size. Particular angular sizes are isolated by varying the scale of the kernel. The wavelet kernel on scale $i$ used in \texttt{wvdecomp} is approximately the difference of two Gaussians, isolating structures in the convolved image of a characteristic scale of $\approx 2^{i-1}$. The input image is convolved with a series of kernels with varying scales, starting with the smallest scale. In each step, significantly detected features of the particular scale are subtracted from the input image before going to the next scale. This allows, among other things, to decompose structures of different sizes into their components, \eg in the case of point-like sources in the vicinity of an extended object. Wavelet kernels have the advantage of a simple linear back transformation, \ie the original image is the sum of the different scales. We define a scale around \SI{0.2}{\rfivehundred} up to which all emission from smaller scales is classified as contamination and is included in our background modelling for the core-modelled single $\beta$-model approach.
The galaxy clusters \SI{0.2}{\rfivehundred} wavelet scales are around \numrange[range-units = single, range-phrase=--]{3}{5} (\numrange[range-units = single, range-phrase=--]{2}{3}) for pointed (survey) observations. This corresponds to \SIrange[range-units = single, range-phrase=--]{4}{16}{\pixel} (\SIrange[range-units = single, range-phrase=--]{2}{4}{\pixel}), with a pixel size of \SI{15}{\arcsec} (\SI{45}{\arcsec}). The detection threshold of a wavelet kernel convolved image is the level above which all maxima are statistically significant. \citet{1998ApJ...502..558V} performed Monte-Carlo simulations of flat Poisson background to define detection thresholds such that one expects on average $1/3$ false detections per scale in a \SI[product-units = single]{512 x 512}{\pixel} image. We adapt a slightly more stringent threshold of $5\sigma$.

\subsubsection{Likelihood function}
\label{sec:likefunc}

Under the assumption that the observed counts are Poisson distributed, the maximum-likelihood estimation statistic to estimate the surface brightness profile parameters is chosen to be the Poisson likelihood.
%
%
The so-called Cash statistic \citep{1979ApJ...228..939C} is derived by taking the logarithm of the Poisson likelihood function and neglecting the constant factorial term of the observed counts
\begin{equation}
\ln\mathcal{L} \propto \sum_{i} O_{i}  \ln\left( M_{i} \right) - M_{i},
\label{equ:lnpoissonlike}
\end{equation}
where $M_{i}$ and $O_{i}$ are the model and observed counts in bin $i$, respectively. The model counts of the background sources using wavelet decomposition, $B_{\textnormal{wv},i}$, are not Poissonian. We assume this background component without error, \ie just the total amount of counts show dispersion. Thus, we can add this background component to the model counts \citep{2016ApJ...827L..38G}.
%
%
In the same way, we add an additional particle background component,  $B_{\textnormal{p},i}$, to equation (\ref{equ:lnpoissonlike}) for pointed observations, \ie the likelihood function becomes
\begin{equation}
\ln\mathcal{L} \propto \sum_{i} O_{i}  \ln\left( M_{i} + B_{\textnormal{wv},i} + B_{\textnormal{p},i} \right) - \left(M_{i}+ B_{\textnormal{wv},i} + B_{\textnormal{p},i} \right).
\label{equ:lnpoissonlikewvbkg+pbkg}
\end{equation}
A single $\beta$- plus constant background model is used to describe the surface brightness of each cluster (see equation (\ref{equ:betamodel}), using $N=1$ and dropping the index $j$)
\begin{equation}
s_{i}(R_{i}) = s_{0} \left[ 1 + \left( \frac{ R_{i} } { r_{\textnormal{c}} } \right)^{2} \right]^{-3 \beta + \num{0.5}} + b_{\textnormal{c}}.
\label{equ:singlebeta+constbkg}
\end{equation}
The projected radii $R_{i}$ are placed at the center of the bins. 
By use of the exposure map, we calculate the proper area, $\alpha_{i}$, and the vignetting corrected mean exposure time, $\epsilon_{\textnormal{mean},i}$. The model counts in each bin are then calculated by multiplying equation (\ref{equ:singlebeta+constbkg}) with the corresponding area and exposure time
\begin{equation}
M_{i} = s_{i}(R_{i}) \cdot \alpha_{i} \cdot \epsilon_{\textnormal{mean},i}.
\label{equ:obscountsmodel}
\end{equation}

\subsection{Point-spread function}

The ability of an X-ray telescope to focus photons, \ie its response to a point source, is characterized by its point-spread function. More peaked cool-core objects are affected more by PSF effects compared to non-cool-core objects. The \rosat PSF depends amongst others on photon energy, off-axis angle and observation mode. A detailed description of the \rosat PSF functions is presented in \citet{2000A&AS..141..507B}. We use the Python package \texttt{pyproffit}\footnote{\url{https://github.com/domeckert/pyproffit}} to calculate PSF mixing matrices based on equations (7) and (30) of \citet{2000A&AS..141..507B} for pointed and survey observations, respectively. These matrices are folded in our surface brightness profile fitting method to obtain PSF unconvolved parameters.

\subsection{Emission measure profiles}

This subsection describes our approach to obtain background subtracted self-similar scaled emission measure profiles. First, the outer significance radius and background level of each galaxy cluster are iteratively determined using the growth curve analysis method \citep{2000ApJS..129..435B,2002ApJ...567..716R}. The outer significance radius determines the maximum radius out to which galaxy cluster emission is detected and thus to which radius each profile is extracted. Background-subtracted and logarithmically binned surface-brightness profiles are converted into emission measure profiles using the normalisation of a partially absorbed Astrophysical Plasma Emission Code (APEC) model
\begin{equation}
\frac{10^{-14}} {4\pi[D_{\textnormal{A}}(1+z)]^{2}} \int n_{\textnormal{e}} n_{\textnormal{H}} \textnormal{d}V.
\label{equ:apecnorm}
\end{equation}
The total weighted hydrogen column density (calculated with the method of \citet{2013MNRAS.431..394W})\footnote{\url{http://www.swift.ac.uk/analysis/nhtot/index.php}} is used to describe the absorption by the atomic and molecular Galactic column density of hydrogen. Metallicities are fixed to \SI{0.35}{\solarmetal} and the abundance table compiled by \citet{1989GeCoA..53..197A} is used. The emission measure along the line-of-sight,
\begin{equation}
EM(R) = \int n_{\textnormal{e}} n_{\textnormal{H}} \textnormal{d}l,
\label{equ:lofemissionmeasure}
\end{equation}
is self-similar scaled according to \citet{2002A&A...389....1A} and 
\begin{equation}
T \propto  \left(E(z) M_{500} \right)^{2/3}
\label{equ:sssTM}
\end{equation}
by
\begin{equation}
\Delta_{\textnormal{z}}^{3/2} (1+z)^{9/2} \left( \frac{ E(z) M_{500}}{\SI[exponent-product = \cdot]{2E15}{\solarmass}} \right)^{1/3},
\label{equ:sssPlanck}
\end{equation}
where $\Delta_{\textnormal{z}}$ is calculated using the density contrast, $\Delta_{\textnormal{c}}$, and matter density parameter at redshift $z$, $\Omega_{\textnormal{z}} = \Omega_{\textnormal{m}} (1+z)^{3} / E(z)^{2}$, according to
\begin{equation}
\Delta_{\textnormal{z}} = \Delta_{\textnormal{c}}
\Omega_{\textnormal{m}} / \left( 18\pi^{2}\Omega_{\textnormal{z}} \right).
\label{equ:deltaz}
\end{equation}
Under the assumption that the cluster has just virialized, \citet{1998ApJ...495...80B} derived an analytical approximation of $\Delta_{\textnormal{c}}$ for a flat universe from the solution to the
collapse of a spherical top-hot perturbation
\begin{equation}
\Delta_{\textnormal{c}} = 18\pi^{2} + 82w - 39w^{2},
\label{equ:deltac}
\end{equation}
with $w = \Omega_{\textnormal{z}} - 1$.

\subsection{Scaling relations}
\label{subsec:scaling_relations_introduction}

In this subsection we describe the basic principle of our linear regression routine to obtain scaling relations. A set of two variates, $x/y$, is fitted by a power-law relation according to
\begin{equation}
\log{y/n_{y}} = m \cdot \log{x/n_{x}} + b.
\label{equ:power_law}
\end{equation}
The pivot elements, $n_{x/y}$, are set to the median along a given axis, such that the results of the slope and normalisation are approximately uncorrelated.

\subsubsection{Likelihood function}

Linear regression of the scaling relations is performed using a Markov chain Monte Carlo (MCMC) posterior sampling technique. We adapt an $N$ dimensional Gaussian likelihood function
\begin{equation}
\mathcal{L}  = \prod_{n=1}^{N} \frac{1}{ 2\pi  \sqrt{\det(\Sigma_n + \Lambda)} }\,
\exp\left(-\frac{1}{2} \tilde{r}_{n}^{\mathrm{\,T}}  (\Sigma_n + \Lambda)^{-1}   \tilde{r}_{n}\right),
\end{equation}
extended compared to \citet{2007ApJ...665.1489K} to account for intrinsic scatter correlation. The intrinsic scatter tensor, $\Lambda$, is described in more detail in  Sect. \ref{sec:covariance}. The uncertainty tensors $\Sigma_n$ account for measurement errors in the independent and the dependent variables and $\tilde{r}_{n}$ denote the residual vectors. For illustration purposes, this is how these two objects would look like in a bivariate example:
\begin{align}
\tilde{r}_n &=  \left(\begin{array}{c}
    x_n - \tilde{x}_n \\
    y_n - m\,\tilde{x}_n - b 
     \end{array}\right) \\
\Sigma_n &= \left(\begin{array}{cc}
    {\sigma_{x,n}}^2 & 0 \\
    0 & {\sigma_{y,n}}^2 \\
    \end{array}\right) 
\end{align}
In this study, the correlation between different measurement errors in the uncertainty tensor is set to zero. The "true" coordinate $\tilde{x}_n$  is normal-scattered according to the intrinsic scatter tensor via
\begin{equation}
\left(\begin{array}{c} \hat{x}_n \\  \hat{y}_n \end{array}\right)  \sim \mathcal{N} \left( \left(\begin{array}{c} \tilde{x}_n \\  m\,\tilde{x}_n - b \end{array}\right), \Lambda \right).
\end{equation}
We integrate out, \ie marginalize over, $\hat{x}_n$ and $\hat{y}_n$. The scatter along the independent axis, $\lambda_{x}$, of the intrinsic scatter tensor is fixed to avoid degeneracies. This means that for this study the intrinsic scatter in temperature is fixed to \SI{20}{\percent}, \ie $\lambda_{\mathrm{T}} = 0.11$ \citep{2006ApJ...650..128K}.
The correlation between the intrinsic scatter values of the two variates $x$ and $y$, $\lambda_{xy}$, is of particular interest for this study and will be described in more detail in Sect. \ref{sec:covariance}. We use the \texttt{emcee} algorithm and implementation \citep{2013PASP..125..306F} for optimization. A chain is considered as converged when the integrated autocorrelation time is greater than one-hundredth of the chain length.  

\subsubsection{Covariance}
\label{sec:covariance}

The linear relationship and thus the joint variability between two or more sets of random variables can be quantified by the covariance between those variates. In the simple case of two variables $x$ and $y$, each with a sample size of $N$ and expected values $\bar{x}$ and $\bar{y}$, the covariance is given by
\begin{equation}
\mathrm{cov}(x,y) \coloneqq \frac{1}{N-1} \sum_{i=0}^N ( x_{i} - \bar{x}_{i} )( y_{i} - \bar{y}_{i} ).
\label{equ:cov}
\end{equation}
The degree of correlation can be calculated by normalizing the covariance to the maximum possible dispersion of the single standard deviations $\lambda_{x}$ and $\lambda_{y}$, the so-called Pearson correlation coefficient:
\begin{equation}
\lambda_{xy} \coloneqq \frac{\mathrm{cov}(x,y)}{\lambda_{x}\lambda_{y}}.
\label{equ:corrcoef}
\end{equation}
The Pearson correlation coefficient can take values between $-1$ and $+1$, where $0$ means no linear correlation and $+1$ ($-1$) means total positive (negative) linear correlation. In the general case of $n$ sets of variables $\{X_{1}\},\dots , \{X_{n}\}$, the covariances can be displayed in a matrix, where the first-order covariance matrix is defined by
\begin{equation}
\Lambda_{l,m} \coloneqq \mathrm{cov}(X_{l},X_{m})  .
\label{equ:covmatrix1}
\end{equation}
In the previous example of 2 variables $x$ and $y$, the covariance matrix reads
\begin{equation}
\Lambda = 
\begin{pmatrix}
\mathrm{cov}(x,x) & \mathrm{cov}(y,x) \\
\mathrm{cov}(x,y) & \mathrm{cov}(y,y)
\end{pmatrix}
=
\begin{pmatrix}
\lambda_{x}^{2} & \lambda_{xy} \lambda_{x} \lambda_{y} \\
\lambda_{xy} \lambda_{x} \lambda_{y} & \lambda_{y}^{2}
\end{pmatrix}
.
\end{equation}
The latter equality makes use of equation (\ref{equ:cov}), which implies that the covariance of a variate with itself, \ie $\mathrm{cov}(x,x)$, reduces to the variance of $x$ or the square of the standard deviation of $x$. The off-axis elements are rewritten by solving equation (\ref{equ:corrcoef}) for the covariance and using the symmetry $\lambda_{xy} = \lambda_{yx}$.
Calculating the Pearson correlation coefficient between the ranked variables is a non-parametric measure of a monotonic relationship between the variables and is called the Spearman rank-order correlation coefficient.

\subsubsection{Selection effects}

As already discussed in Sect. \ref{sec:introduction}, centrally peaked galaxy clusters are more likely to enter an X-ray selected sample due to their enhanced central emission. \citet{2011A&A...532A.133M} investigated this effect by applying the \hiflugcs flux limit to Monte Carlo simulated samples. Assuming \hiflugcs being complete, one can vary the input fractions of different core-types in the simulations to match the observed ones. The intrinsic scatter increases the normalization of the luminosity-temperature relation because up-scattered clusters have a higher chance of lying above the flux threshold. 
In this study, we are not trying to determine the true luminosity-temperature relation but are interested in the residuals of the sample with respect to the mean to study the intrinsic scatter covariances. Therefore we are neglecting Malmquist bias in the parameter optimization, although it is present in \hiflugcs.
To investigate the effect of Malmquist bias on the best-fit shape-temperature relation parameters and the intrinsic scatter correlation coefficients, we artificially decrease the luminosity-temperature relation normalization and find that the differences are insignificant.

\subsection{Cool-core classification}

\citet{2010A&A...513A..37H} used \hiflugcs to compare $16$ different techniques to differentiate cool-core and non-cool-core clusters. The central cooling time, $t_{\textnormal{cool}}$, was found to be suited best and used to classify clusters into three categories. Clusters with central cooling times shorter than \SI{1}{\giga\yr} are classified as
strong-cool-core (SCC) clusters. They usually show characteristic temperature drops towards the center and low central entropies. Clusters exhibiting high central entropies and cooling times greater than \SI{7.7}{\giga\yr} are classified as non-cool-core (NCC) clusters. In intermediate class with cooling times in between those of SCC and NCC clusters are classified as weak-cool-core (WCC) clusters. We adapt the \citet{2010A&A...513A..37H} classification scheme and categorization of \hiflugcs clusters for this study. There are \numcolor{\num{45}} galaxy clusters above our selected temperature threshold of \SI{3}{\keV} with mass estimates in the Planck "Union catalogue". The amount of each core-type category is \numcolor{\num{15}}, \numcolor{\num{16}} and \numcolor{\num{14}} for SCC, WCC and NCC, respectively. For \num{1} of the SCC and \num{3} of the WCC objects, no \rosat pointed observations are available and \rass data is used.

%
\section{Emission measure profiles}
\label{sec:emprofiles}

\begin{figure*}
\centering
\resizebox{\hsize}{!}{\includegraphics{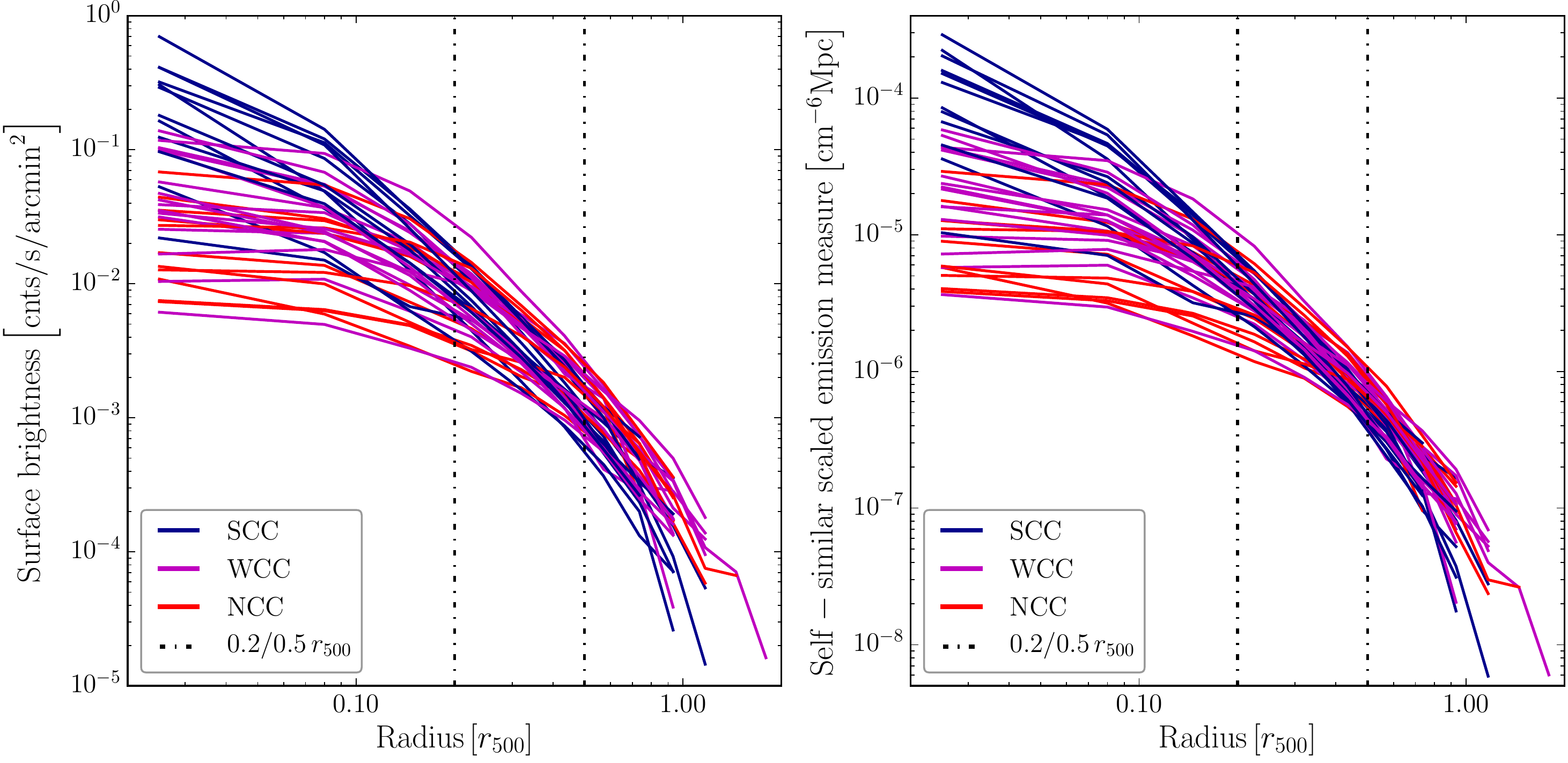}}
\caption{Surface brightness  (\textit{left}) and self-similar scaled emission measure profiles (\textit{right}) for \hiflugcs objects with temperatures greater than \SI{3}{\keV}.
}
\label{fig:sb_sscEM_kTbin}
\end{figure*}
\begin{figure*}
\centering
\resizebox{\hsize}{!}{\includegraphics{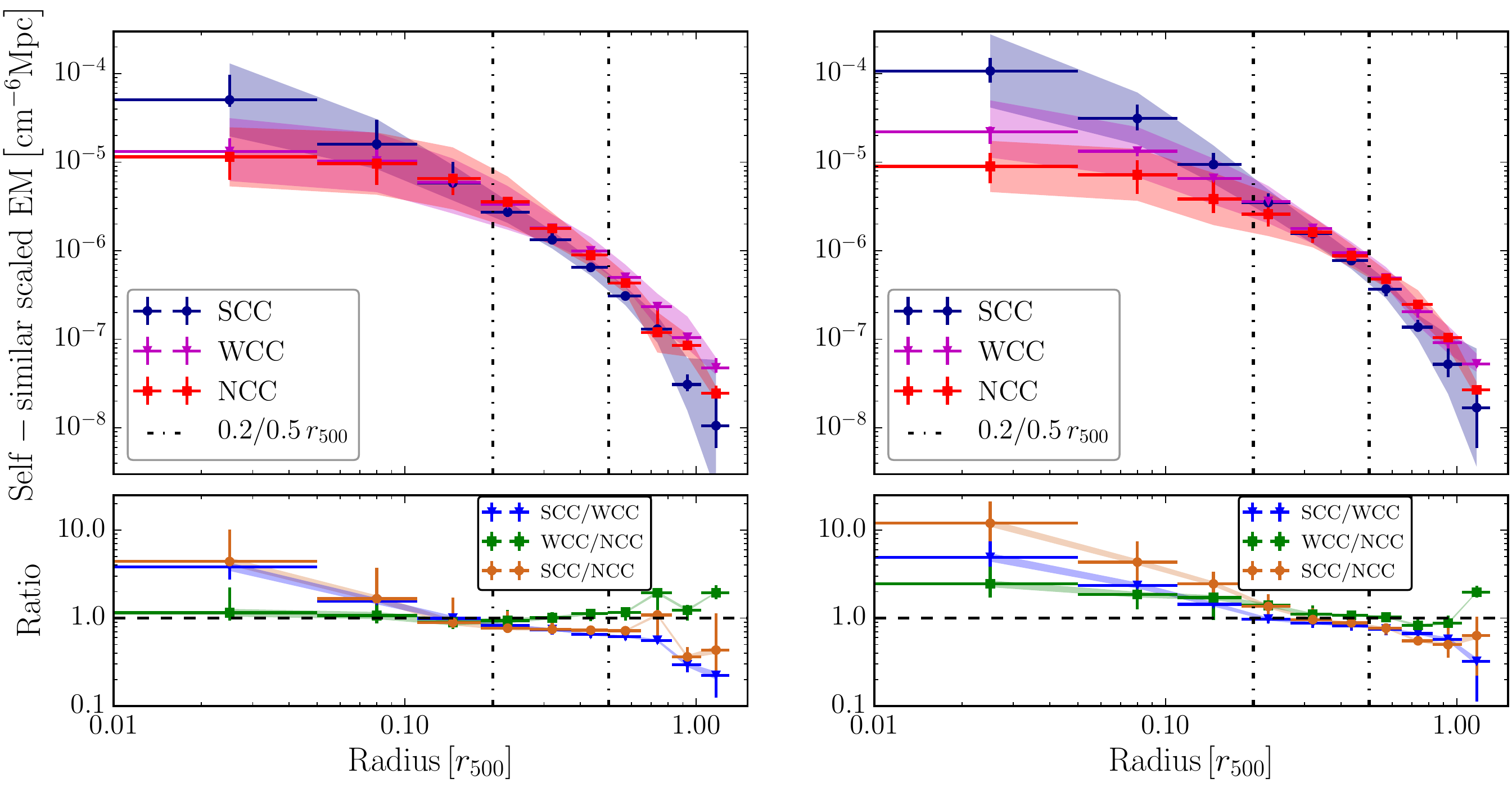}}
\caption{\textit{Upper panels}: Weighted mean (\textit{left}) and median (\textit{right}) self-similar scaled emission measure profiles for all \hiflugcs objects with temperatures greater than \SI{3}{\keV} and for the individual core types. \textit{Lower panels}: Ratio of the self-similar scaled emission measure profiles between the different core type populations. Shown error bars were estimated with \num{10000} bootstrapping iterations. The shaded regions represent the intrinsic scatter values of each bin.
}
\label{fig:sscEM_stacked_kTbin}
\end{figure*}

This section describes a model-independent way to test the surface brightness (SB) behaviour between different core-type populations by comparing their background subtracted self-similar scaled emission measure (EM) profiles. Fig. \ref{fig:sb_sscEM_kTbin} shows the SB and self-similar scaled EM profiles. Outside the cluster core the galaxy clusters show self-similar behaviour, as already discussed in several previous studies (\eg \citealt{1999ApJ...525...47V}). The self-similar scaled EM profiles show a smaller intrinsic scatter compared to the surface brightness profiles in the \SIrange[range-units = single, range-phrase=--]{0.2}{0.5}{\rfivehundred} range. It is reduced by \numcolor{$28$\%} (\numcolor{$33$\%}) with respect to the median (weighted mean).
To investigate possible differences between the core-type populations, the emission measure profiles are stacked according to their core types by calculating the weighted mean and the median, as shown in the upper panel of Fig. \ref{fig:sscEM_stacked_kTbin}. The lower panel of Fig. \ref{fig:sscEM_stacked_kTbin} shows the ratio of the average EM profiles sorted according to their core types. The statistical errors on the weighted means or medians are very small. The error bars indicated on the plot correspond to uncertainties calculated by bootstrapping the data, \ie by shuffling the profiles with repetition, repeating the operation \num{10000} times and computing the median and percentiles of the output distribution. The bootstrap errors thus include information on the sample variance and non-Gaussianity of the underlying distribution.

\subsection{Discussion}
\label{subsec:emprofiles_discussion}

The weighted mean profiles reveal in a model-independent way the existence of subtle differences between the galaxy cluster populations. The amplitude of this effect between SCC and NCC clusters in the \SIrange[range-units = single, range-phrase=--]{0.2}{0.5}{\rfivehundred} radial range is up to $30$\% and confirms the finding of \citet{2012A&A...541A..57E}. Compared to the heterogeneous sample of \citet{2012A&A...541A..57E} \hiflugcs is statistically complete, which confirms the result in a more robust way. If true, this finding implies that the outskirts are affected by the core-type and a detection algorithm tailored to the galaxy cluster outskirts will be more sensitive to the more abundant NCC objects, which needs to be taken into account in selection functions. In addition one could determine the statistical likelihood for the core-state of a particular galaxy cluster.
However, the asymmetric bootstrap errors indicate an underlying non-Gaussian distribution or that the sample is affected by outliers. The median profiles, which are more robust against outliers, do not reveal the same trend of the emission measure ratios. As a test, we exclude the strong cool-core profile that deviates most with respect to the median within \SIrange[range-units = single, range-phrase=--]{0.2}{0.5}{\rfivehundred} (Abell 3526). This cluster also shows the smallest statistical errors in this radial range. The weighted mean ratios resembles well the trend of the median when excluding this single cluster from the analysis. This reveals that the weighted mean is driven by an outlier with small statistical errors. Therefore, we conclude that there is no indication of a systematic core-type differences in the galaxy clusters \SIrange[range-units = single, range-phrase=--]{0.2}{0.5}{\rfivehundred} radial range. Nevertheless, this comparison is useful because outliers like Abell 3526 will affect the selection function.
The investigation of a possible redshift evolution of this analysis is left to a future study. Beyond \SI{0.5}{\rfivehundred} the difference between SCC and NCC clusters become larger. \citet{2012A&A...541A..57E} discussed gas redistribution between the core region and the outskirts as possible explanation. In this scenario, the injected energy due to a merging event flattens the density profile of interacting objects. Assuming that NCC clusters are more likely to have experienced a recent merging event, their self-similar scaled EM profiles would be different compared to CC objects. Another explanation could be the current accretion of large scale blobs. Clusters with higher mass accretion rates show a larger fraction of non-thermal pressure in simulations \citep{2014ApJ...792...25N}. Again assuming that NCC objects are merging clusters, the discrepancies can be explained by the different non-thermal energy content. However, this scenario seems unlikely since we expect to detect such structures in our wavelet images. An additional explanation can be that the dark matter halos of NCC and CC objects have different shapes and thus a different concentration at a given radius.

%

%
\section{Large scale center and ellipticity}

\begin{figure}
\centering
\resizebox{\hsize}{!}{\includegraphics{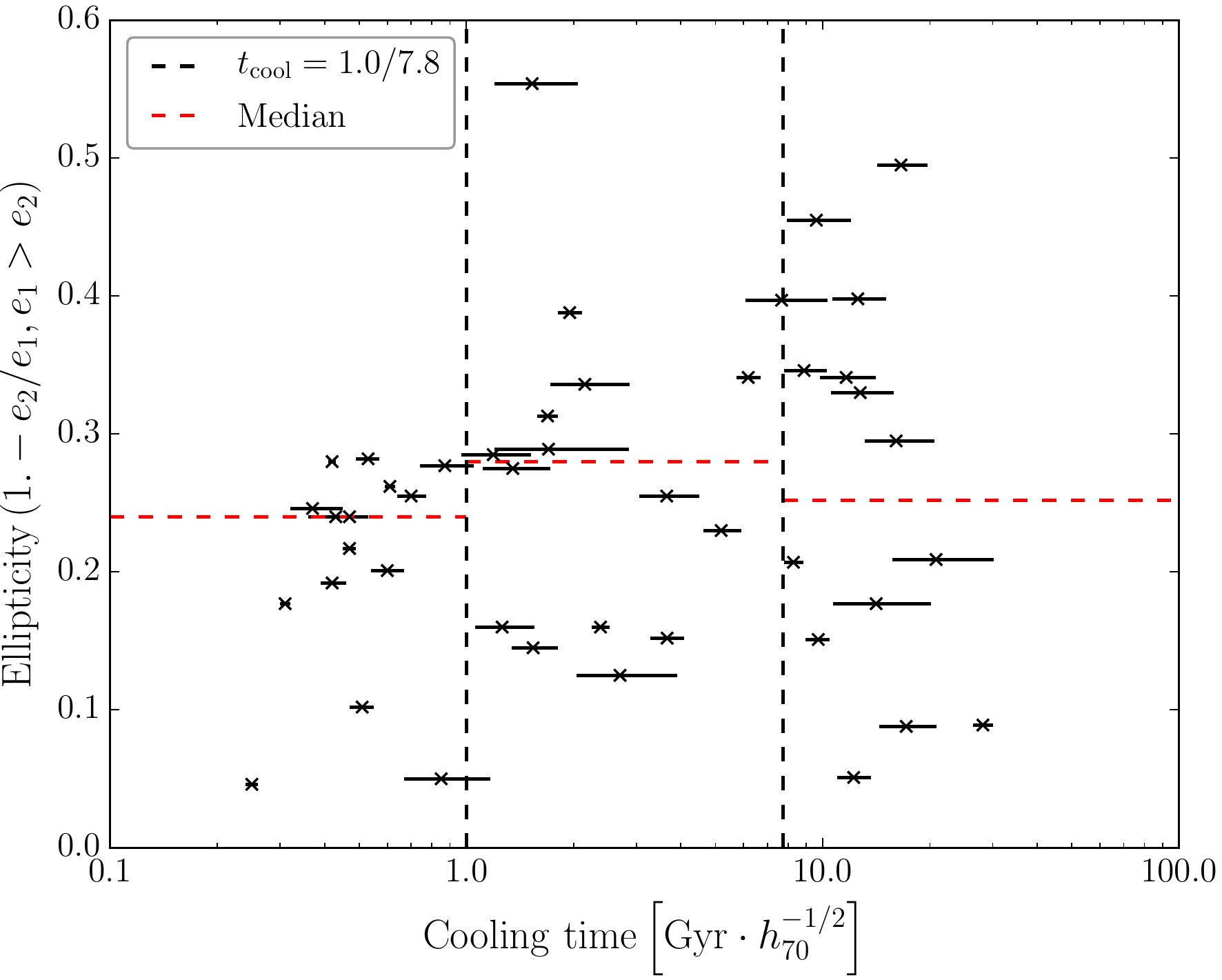}}
\caption{Ellipticity as a function of cooling time of \hiflugcs objects with temperatures greater than \SI{3}{\keV}.
}
\label{fig:cct_vs_ellipticity}
\end{figure}

The wavelet decomposition allows us to study galaxy cluster parameters on large scales. The one scale around \SI{0.2}{\rfivehundred} is used to determine the center and ellipticities using the \texttt{SExtractor} program \citep{1996A&AS..117..393B}. This minimizes the impact of the different core states on these parameters. The chosen scale contains most of the cluster counts outside the core region and is therefore a good tracer for large scale properties. Including scales up to around \SI{0.5}{\rfivehundred} (an additional \numrange[range-units = single, range-phrase=--]{1}{3} scales) shows that the median and mean difference in ellipticity is just about 10\%. The ellipticity, $e$, of an ellipse with major-axis, $e_{1}$, and minor-axis, $e_{2}$, is defined as 
\begin{equation}
e = 1 - e_{2}/e_{1}
\label{equ:ellipticity}
\end{equation}
and is shown as function of the cooling time in Fig. \ref{fig:cct_vs_ellipticity}. The two parameters do not show a significant correlation (Spearman rank-order correlation coefficient of \numcolor{\num{0.17}}). The medians and sample standard deviations in the 3 bins are (\numcolor{\num{0.24}}, \numcolor{\num{0.28}}, \numcolor{\num{0.25}}) and (\numcolor{\num{0.08}}, \numcolor{\num{0.12}}, \numcolor{\num{0.14}}), respectively. There are no highly elliptical clusters with short cooling times, \ie selecting clusters above an ellipticity of $\sim$\num{0.3} creates a sample without SCC objects. The universality of this result needs to be confirmed with galaxy cluster samples of larger sizes.

\begin{figure*}
\centering
\resizebox{\hsize}{!}{\includegraphics{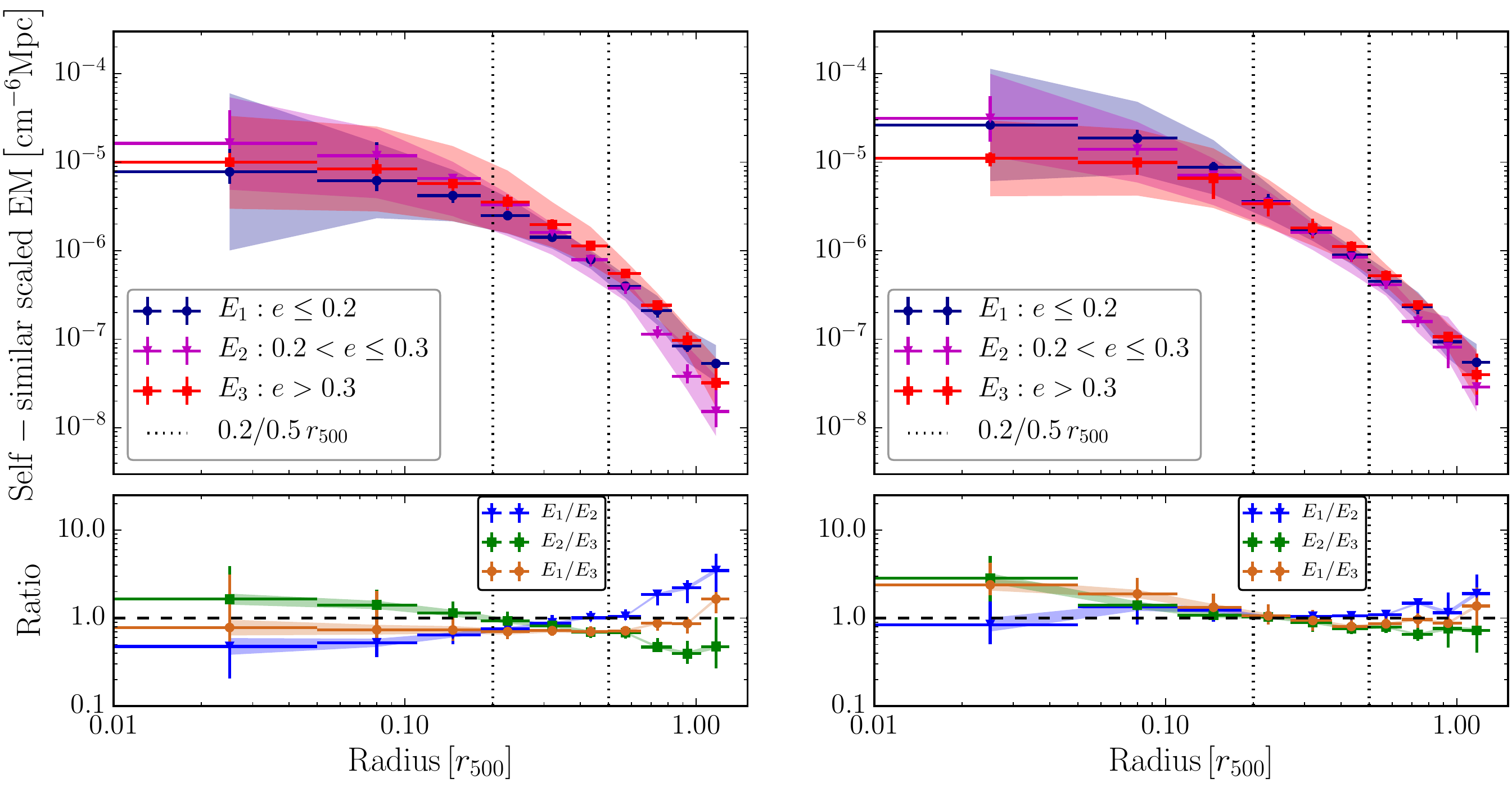}}
\caption{Same as Fig. \ref{fig:sscEM_stacked_kTbin}, except that the classification is done according to ellipticity.
}
\label{fig:sscEM_stacked_kTbin_ellipticity}
\end{figure*}

We perform a similar stacking analysis as in Sect. \ref{sec:emprofiles}, where the sample is divided into 3 sub-classes according to ellipticity, rather than core-state. The weighted mean and median profiles are shown in Fig. \ref{fig:sscEM_stacked_kTbin_ellipticity}. The median profiles do not show a difference between the sub-classes except in the core and the very outskirts.

We quantify the covariance between ellipticity and core radius in kpc. For a core-modelled single $\beta$-model and fixed $\beta$-parameter, the Spearman rank-order correlation coefficient is \num{0.20}, \ie no strong indication for a correlation. Allowing the $\beta$-parameter to vary introduces a small positive correlation coefficient of \num{0.33}. A similar behaviour is seen for best-fit core radii using a single $\beta$-model.

\subsection{Discussion}

Clusters with large cooling times show a large range of 2D ellipticities. Assuming that they originate from the same population in 3D, the large scatter of ellipticity might be explained by projection effects, \ie triaxial halos with random orientations will yield a wide range of observed 2D ellipticities. The verification of this effect using simulations is left to a future study. Considering ellipticity adds supplementary information since it is not significantly correlated with the core radius. The ellipticity in the galaxy cluster outskirts traces the amount of baryon dissipation \citep{2012ApJ...755..116L}. Thus, ellipticity is linked and may help to constrain cosmological parameters like the amplitude of the matter power spectrum ($\sigma_{8}$). Halos form later for lower $\sigma_{8}$ values and are therefore on average more elliptical \citep{2006MNRAS.367.1781A,2008MNRAS.391.1940M}. In addition, measuring ellipticity on several scales allows to indirectly study large-scale gas rotation in galaxy clusters \citep{2013MNRAS.434.1565B}. This allows us to test if the ICM is in hydrostatic equilibrium, an often made assumption in X-ray mass measurements. This makes the ellipticity an interesting survey measure for the future \erosita X-ray all-sky survey, where we expect a range of ellipticities.
The ratio of the weighted mean profiles between objects with low and high ellipticity shows a slight offset. Since this offset is not visible in the median profile, it can't solely be explained by the larger non-Gaussianity per bin of elliptical surface brightness distributions extracted in circular annuli.

\citet{2007A&A...467..485H} studied the relationships between the X-ray morphology and several other cluster properties using a heterogeneous sample of \num{101} clusters taken from the \chandra archive, out of which \num{18} objects are represented in \hiflugcs. The ellipticity measurements are in agreement with our study within a factor of $\sim$2, except for A0399. For this cluster, the ellipticity calculated by \citet{2007A&A...467..485H} is a factor of \num{5.6} larger (\num{0.284} compared to \num{0.051}). We note that the semi-major and semi-minor axis are calculated in the same way, \ie by the 2nd-order moments but we subtracted the central excess emission of the cluster before. Including the central excess emission brings our results in better agreement with \citet{2007A&A...467..485H}, especially in the case of A0399. For a single $\beta$-model description of the surface brightness profile, they find a slightly larger Spearman rank-order correlation coefficient of \num{0.37} (compared to \num{0.26}) between ellipticity and core radius in kpc.

%
\section{Analysis of the residuals}
\label{sec:analysis of the residuals}

In this section we quantify how well the different model parameterisations represent the underlying \hiflugcs surface brightness profiles. All profiles are fitted over the full extracted radial range (see Fig. D.1) using an MCMC posterior sampling technique, taking uncertainties in the measured surface brightness into account. 
In case of the core-modelled single $\beta$-model all contaminating sources, including the excess emission in the cluster core, are modelled according to equation (\ref{equ:lnpoissonlikewvbkg+pbkg}). For the single $\beta$-model fitting procedure we perform a classical approach of masking contaminating sources that are detected on the wavelet images. This includes point-like sources or extended emission like substructure but not emission from the cluster core. Thus the likelihood functions for the two cases are different because for a single $\beta$-model, $B_{\textnormal{wv},i}$ in equation (\ref{equ:lnpoissonlikewvbkg+pbkg}) is equal to zero for all $i$.
The choice of priors is discussed in appendix \ref{sec:priors}.
The background is modelled with an additive constant. We note that not subtracting the background before fitting introduces a positive degeneracy between the best-fit slope of the $\beta$-model and the background level. In this framework, residuals are defined as (data-model)/model, \ie positive (negative) residuals indicate that the model under- (over-) predicts the observed surface brightnesses. We focus particularly on the \SIrange[range-units = single, range-phrase=--]{0.2}{0.5}{\rfivehundred} radial range because most of the galaxy cluster counts outside the non-scalable core regions are expected there.

\subsection{Discussion: Single $\beta$-models}
\label{subsec:single_beta_models}

A single $\beta$-model is a widely used description to fit surface brightness profiles, especially in the low statistics regime. Therefore it is also commonly used to detect extended sources in X-ray surveys. Fig. \ref{fig:residuals_from_beta_model_0205r500} (\textit{upper left panel}) shows the fractional median of the residuals from a single $\beta$-model in the \SIrange[range-units = single, range-phrase=--]{0.2}{0.5}{\rfivehundred} radial range as function of core radius. There is a clear trend of positive residuals towards smaller core radii. This means that the flux in the outskirts is systematically under-predicted, especially for SCC objects (see Sect. \ref{subsec:fluxcomp}). In addition, the core radii of objects that exhibit cool-cores are systematically biased low since a single $\beta$-model lacks degrees of freedom to model the central excess emission. Due to higher photon statistics in the core region, a single $\beta$-model fit tends to be driven by the inner radial bins. As an additional test we tried to reduce the weight of the core region in the fitting procedure by assuming that the variance in the Gaussian likelihood function is underestimated by a given fractional amount $f = 0.1$. The qualitative behaviour of the residuals remains the same. In cases where the cluster outskirts are poorly fitted, the background level is not determined properly as well because its level compensates for the poor fit in the outskirts. This influences the $\beta$ value determination as already discussed. The median of the $\beta$ parameters is \num{0.59} and thus smaller than the often assumed generic value of $2/3$. The best-fit values of the surface brightness parameters are in good agreement with \citet{2002ApJ...567..716R}.
The single $\beta$-model residuals for individual clusters have a wavy form. As additional test we investigated different functional forms to describe the common deviations from a single $\beta$-model, but no significant common form of the residuals could be found. The scatter around $0$ and bias in the \SIrange[range-units = single, range-phrase=--]{0.2}{0.5}{\rfivehundred} radial bin are \num{0.092} and $0.032 \pm 0.003$, respectively. The non-zero bias reflects the same finding as discussed above.
The residuals in the \SIrange[range-units = single, range-phrase=--]{0.2}{0.5}{\rfivehundred} radial range from a single $\beta$-model with fixed $\beta$ parameter ($\beta=2/3$) is shown in the \textit{lower left panel} of Fig. \ref{fig:residuals_from_beta_model_0205r500}. The amplitude of the residuals for SCC objects increases up to over \numcolor{40}\%. Fixing the $\beta$ parameter increases the scatter and bias to \numcolor{$0.161$}  and \numcolor{$0.066 \pm 0.003$}, respectively.

\subsection{Discussion: Core-modelled single $\beta$-models}

\begin{figure*}
\begin{subfigure}{0.475\textwidth}
\includegraphics[width=\textwidth]{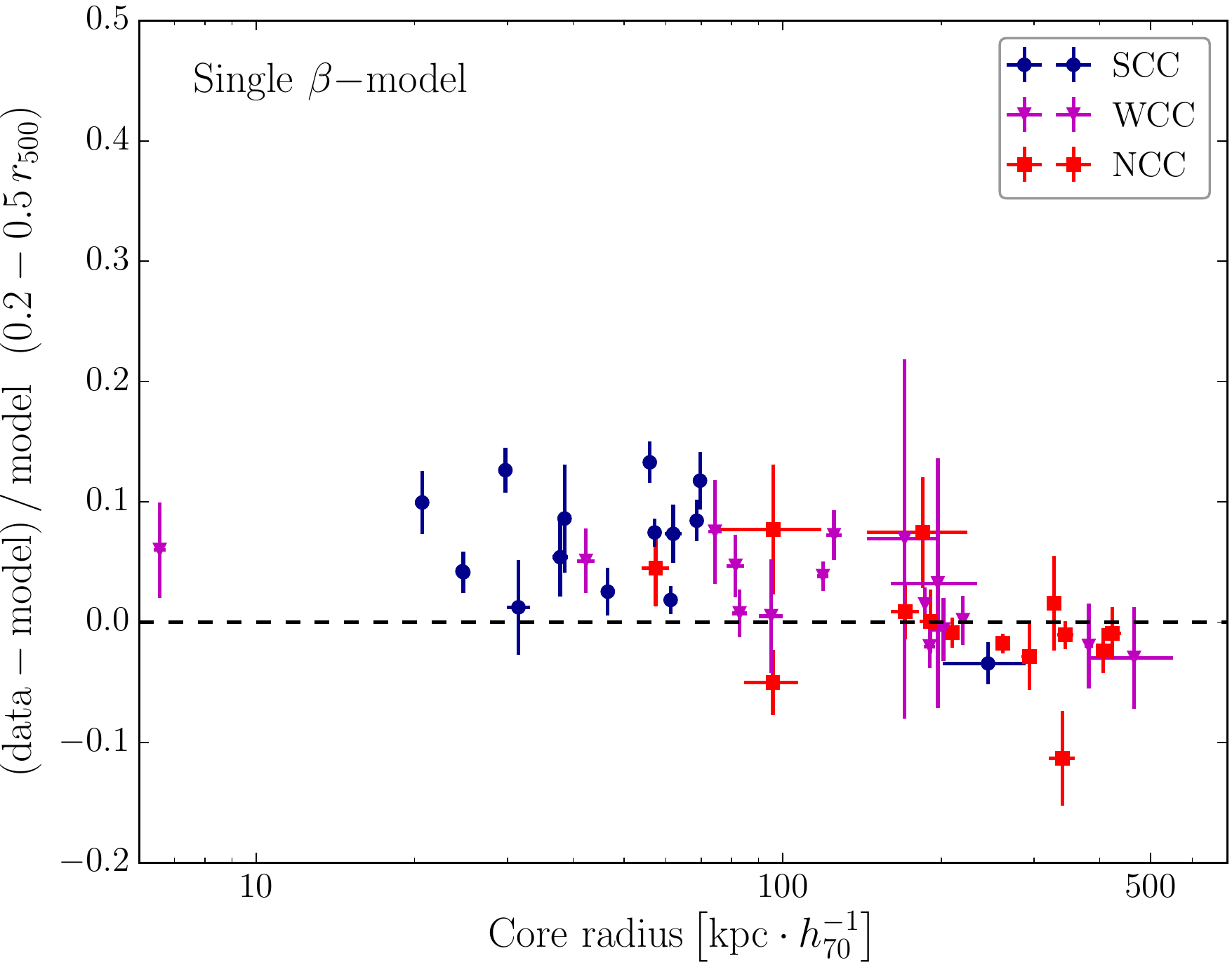}
\label{fig:all_single_beta_bkg_no_scatter_logbin_psexcluded_rc_res_02_05_unstacked}
\end{subfigure} \hspace{0.05\textwidth}
\begin{subfigure}{0.475\textwidth}
\includegraphics[width=\textwidth]{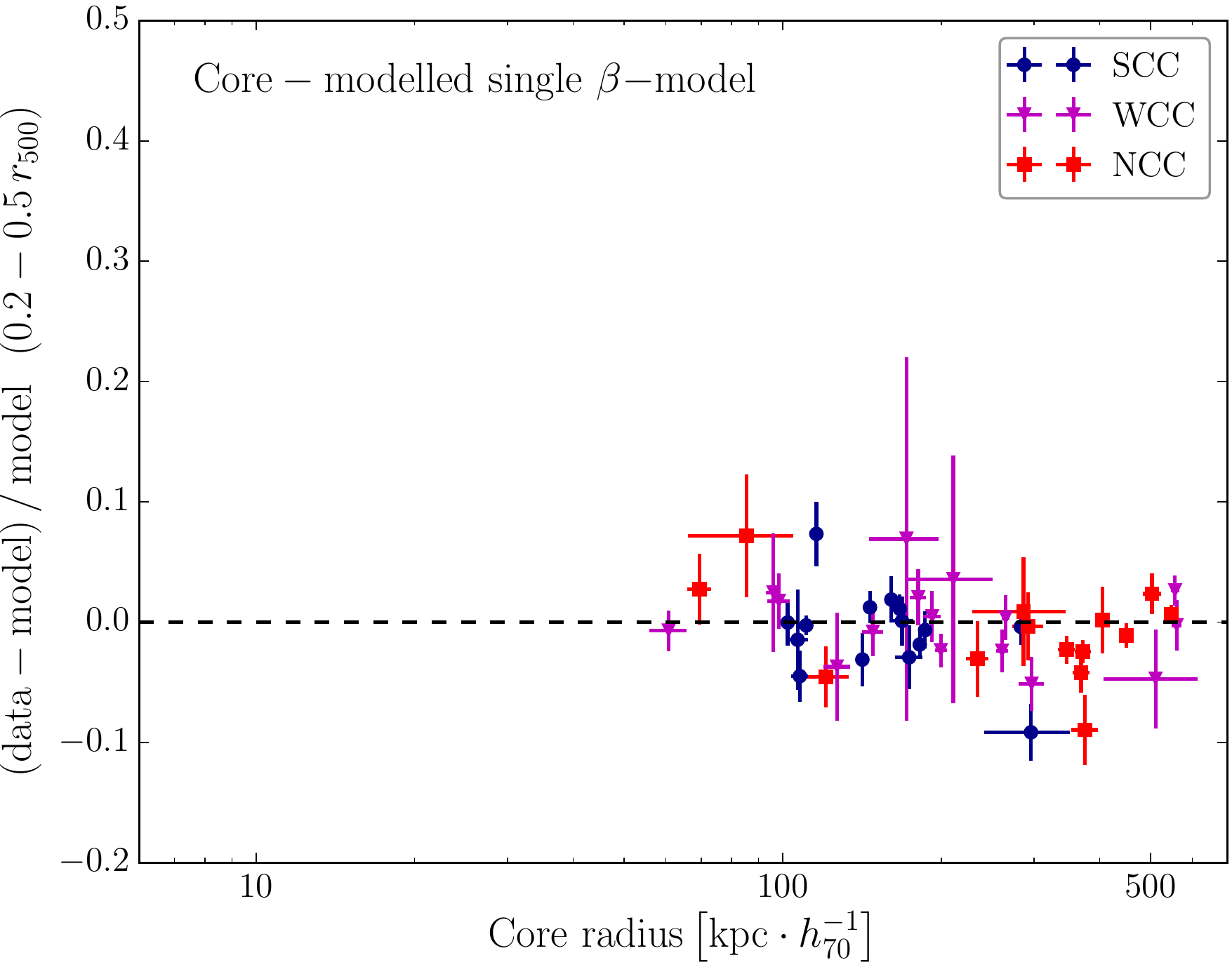}
\label{fig:all_single_beta_bkg_no_scatter_logbin_rc_res_02_05}
\end{subfigure}
\begin{subfigure}{0.475\textwidth}
\includegraphics[width=\textwidth]{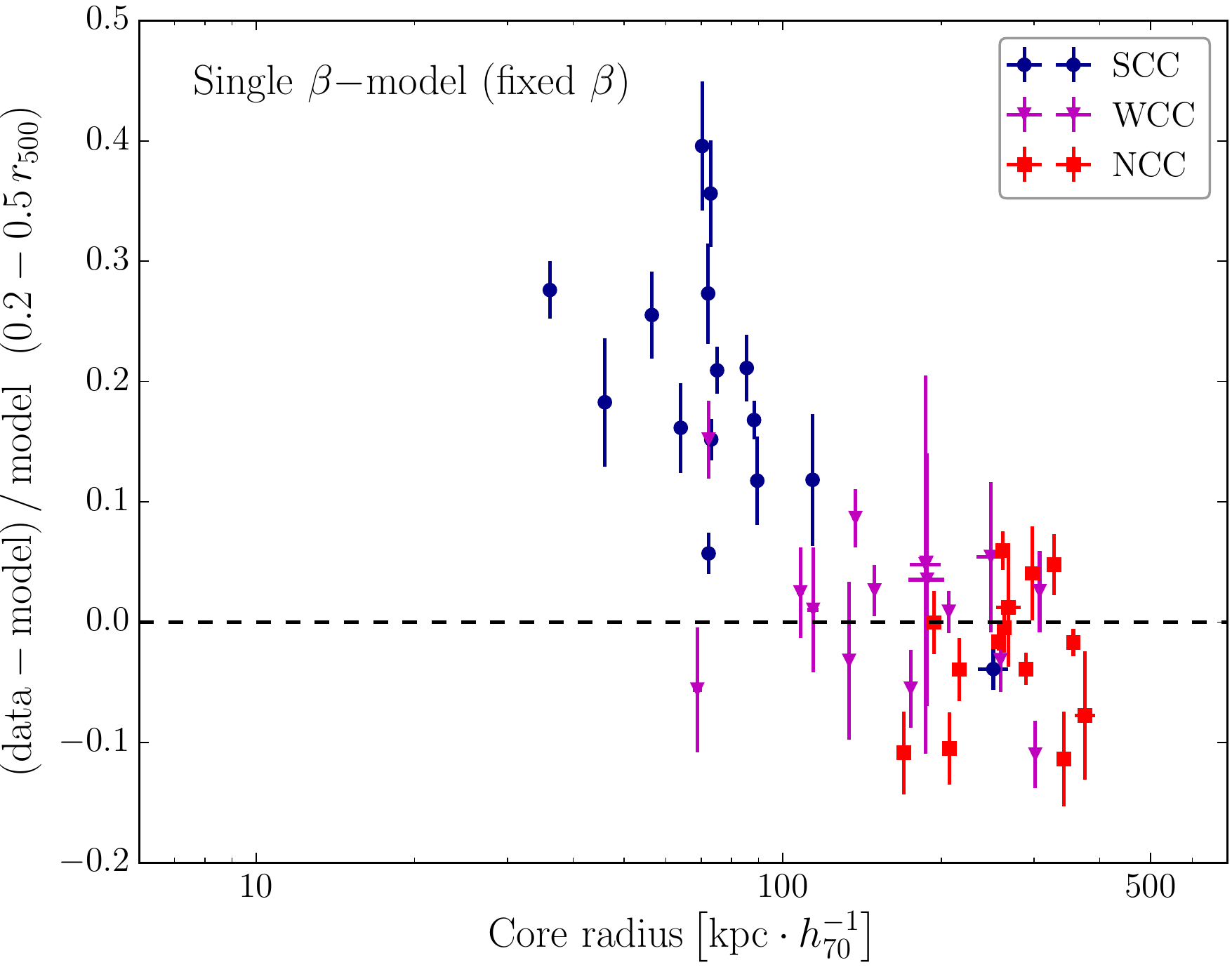}
\end{subfigure}
\hspace{0.05\textwidth}
\begin{subfigure}{0.475\textwidth}
\includegraphics[width=\textwidth]{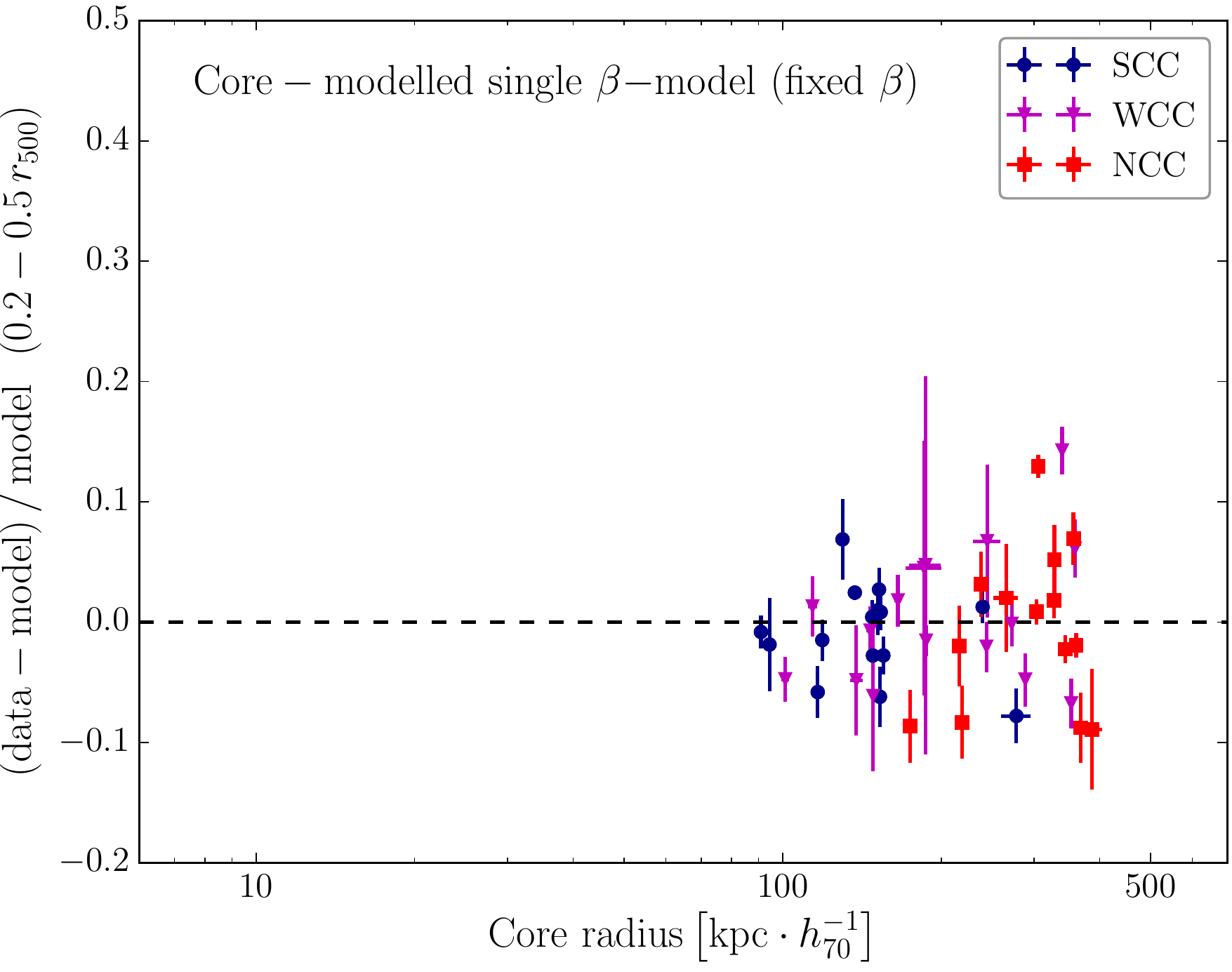}
\label{fig:all_residuen_unstacked}
\end{subfigure}
\caption{
Median of the fractional residuals from a single $\beta$-model (\textit{left panel}) and a core-modelled (\textit{right panel}) single $\beta$-model in \SIrange[range-units = single, range-phrase=--]{0.2}{0.5}{\rfivehundred} radial bins for individual \hiflugcs objects above a temperature of \SI{3}{\keV} as function of core radius. For the \textit{lower panel} the $\beta$ parameter is fixed to $2/3$ in the surface brightness parameter optimization.
}
\label{fig:residuals_from_beta_model_0205r500}
\end{figure*}

The negative effects of a single $\beta$-model as described in Sect. \ref{subsec:single_beta_models} are reduced when modelling the excess core emission by adding the counts on the small scales of the wavelet decomposition to the model counts in the Poisson likelihood function (see Sect. \ref{sec:likefunc}). The residuals in the \SIrange[range-units = single, range-phrase=--]{0.2}{0.5}{\rfivehundred} radial range are shown in Fig. \ref{fig:residuals_from_beta_model_0205r500} (\textit{upper right panel}). The measured residuals scatter around zero and there is no bias between individual core types visible. The scatter around $0$ in the \SIrange[range-units = single, range-phrase=--]{0.2}{0.5}{\rfivehundred} radial bin is slightly reduced compared to the single $\beta$-model (\numcolor{$0.074$}). Most importantly, the bias gets more consistent with zero (\numcolor{$-0.004 \pm 0.003$}). In case of a fixed $\beta$ parameter (\textit{lower right panel} of Fig. \ref{fig:residuals_from_beta_model_0205r500}), the bias of a core-modelled single $\beta$-model is consistent with zero (\numcolor{$-0.001 \pm 0.003$}) and the scatter is slightly increased (\numcolor{$0.087$}). The median of the $\beta$ parameters is larger compared to the single $\beta$-model case and with \numcolor{\num{0.696}} close to generic value of $2/3$.

To verify our core-modelling method and to confirm its results we excise the core region ($<\SI{0.2}{\rfivehundred}$) in the single $\beta$-model fitting procedure. This is an independent way to avoid the single $\beta$-model fit to be driven by the non-scalable core emission and to reduce residuals between model and data in the cluster outskirts. In more than \numcolor{90}\% of the cases, we find that this approach delivers comparable best-fit parameters for $\beta$ and core-radii as when modelling the core emission. Due to excluding data, the constraining power is reduced and the degeneracy of the shape parameters with the $\beta$-model normalization is larger. In several cases, this results in a larger mismatch of the $\beta$-model fluxes compared to the real flux of the objects.

\subsection{Flux comparison}
\label{subsec:fluxcomp}

This section compares the overall model flux, as well as the model flux in the cluster outskirts for the single $\beta$-model and core-modelled single $\beta$-model 
to the measured flux in the corresponding radial range. For each model, we calculate the cluster count rate by integrating a single $\beta$-model with the corresponding best-fit parameters in a given radial range, \ie $0$$-$$r_{\textnormal{x}}$ (the outer significance radius) for the overall model flux and \SIrange[range-units = single, range-phrase=--]{0.2}{0.5}{\rfivehundred} for the flux in the outskirts. We are interested in flux ratios, in which the count rate to flux conversion factors of individual clusters cancel each other out.
%
We calculate the total flux of the single $\beta$-model including the central excess emission and subtract the wavelet-detected central excess emission in the core-modelled case. The median of the measured fluxes in the outskirts and the overall fluxes of the NCC and WCC objects agree very well with each other. The measured overall single $\beta$-model fluxes of the SCC objects are on average $\sim\numcolor{23}$\% larger compared to the measured overall core-modelled fluxes and the flux ratio has an intrinsic scatter of $\sim\numcolor{14}$\% around the median.
The median accuracy between model and measured total flux is within $\sim \numcolor{4}$\% ($\sim\numcolor{2}$\%) for the (core-modelled) single $\beta$-model, regardless of the core-type or if the $\beta$ value is free to vary or fixed to $2/3$. The flux in the outskirts of the core-modelled single $\beta$-model has the same median accuracy of $\sim\numcolor{2}$\%. For the single $\beta$-model, the accuracy stays at the $\sim\numcolor{4}$\% level for WCC and NCC objects. For SCC objects, for the single beta-model case, the flux in the outskirts is biased low by $\sim\numcolor{6}$\% and increases to $\sim\numcolor{10}$\% when $\beta$ is fixed to $2/3$; \ie biased \numrange[range-units = single, range-phrase=--]{3}{5} times larger than the for the core-modelled case. In all cases, the intrinsic scatter values of the ratios are below $\sim\numcolor{6}$\%. The Spearman rank-order correlation coefficients do not reveal a significant correlation between the flux underestimation and the cluster temperature.

%
\section{Scaling relations}

\begin{figure*}
\begin{subfigure}{0.475\textwidth}
\includegraphics[width=\textwidth]{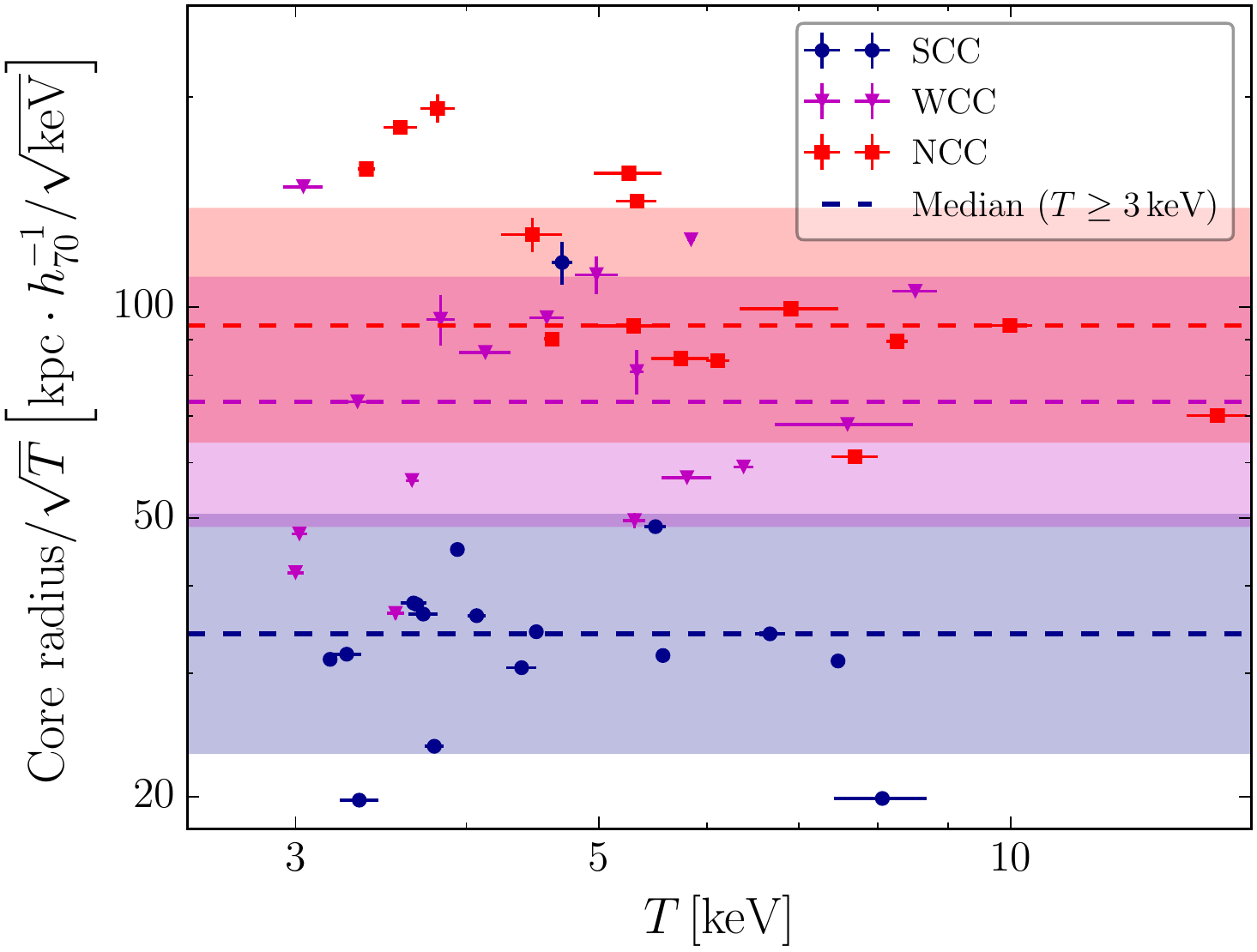}
\end{subfigure} \hspace{0.05\textwidth}
\begin{subfigure}{0.475\textwidth}
\includegraphics[width=\textwidth]{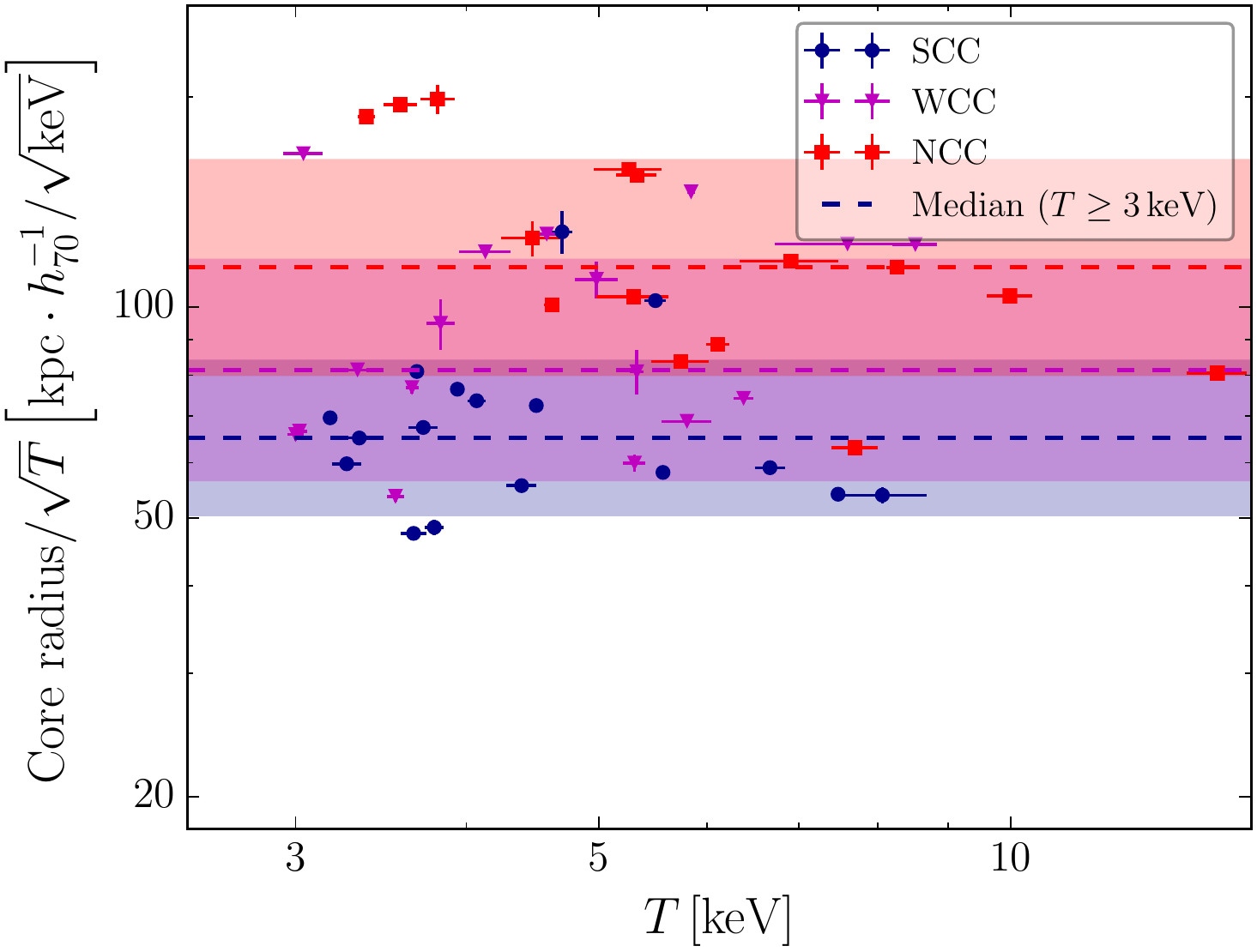}
\end{subfigure}
\caption{
Core radius as a function of temperature for a single $\beta$-model (\textit{left}) and core-modelled (\textit{right}) single $\beta$-model. The dashed lines and shaded regions represent the medians and the their intrinsic scatter of the individual core populations.}
\label{fig:rc_kT_fixed_beta_logbin}
\end{figure*}

\begin{figure*}
\begin{subfigure}{0.475\textwidth}
\includegraphics[width=\textwidth]{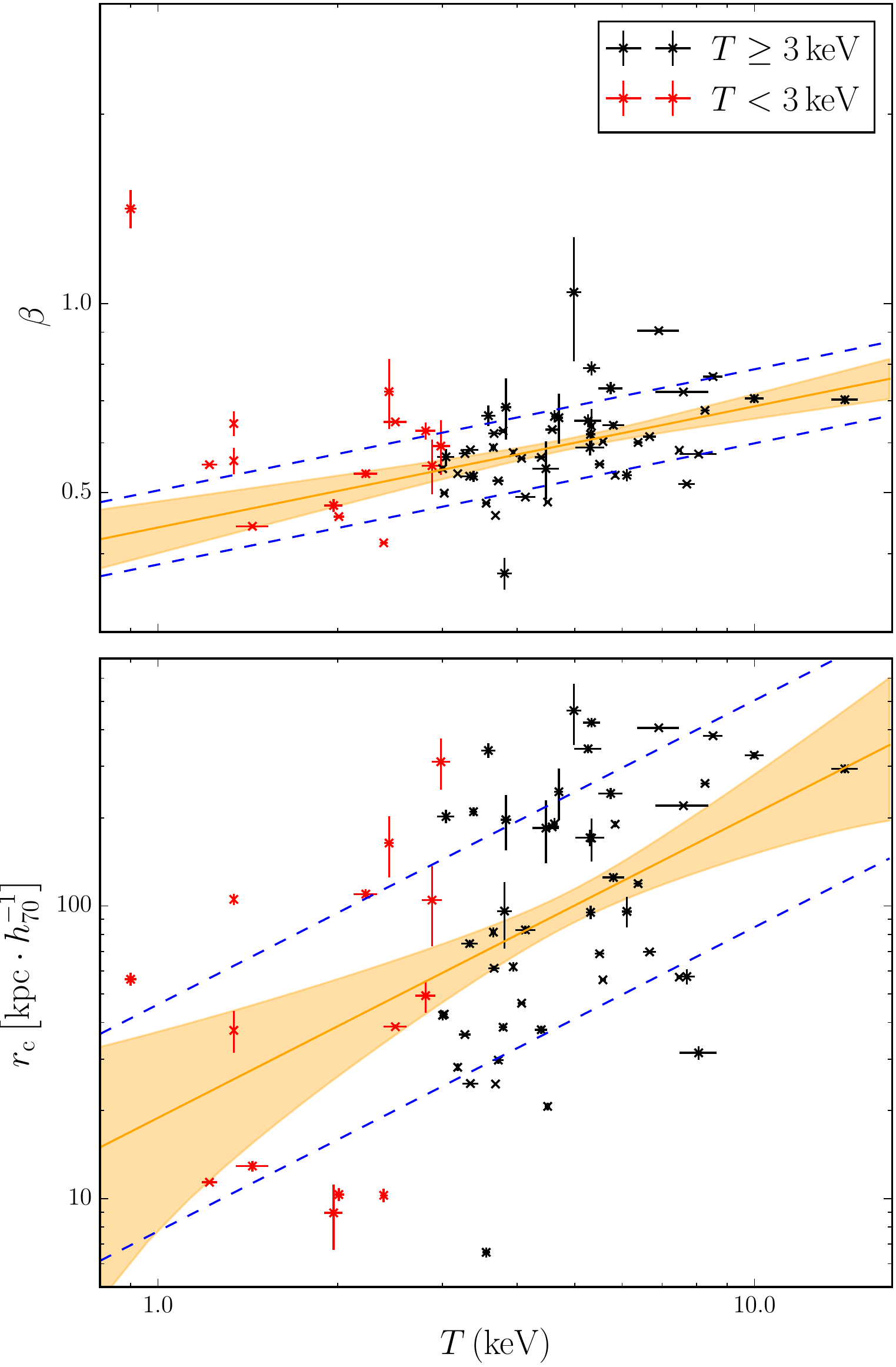}
\end{subfigure} \hspace{0.05\textwidth}
\begin{subfigure}{0.475\textwidth}
\includegraphics[width=\textwidth]{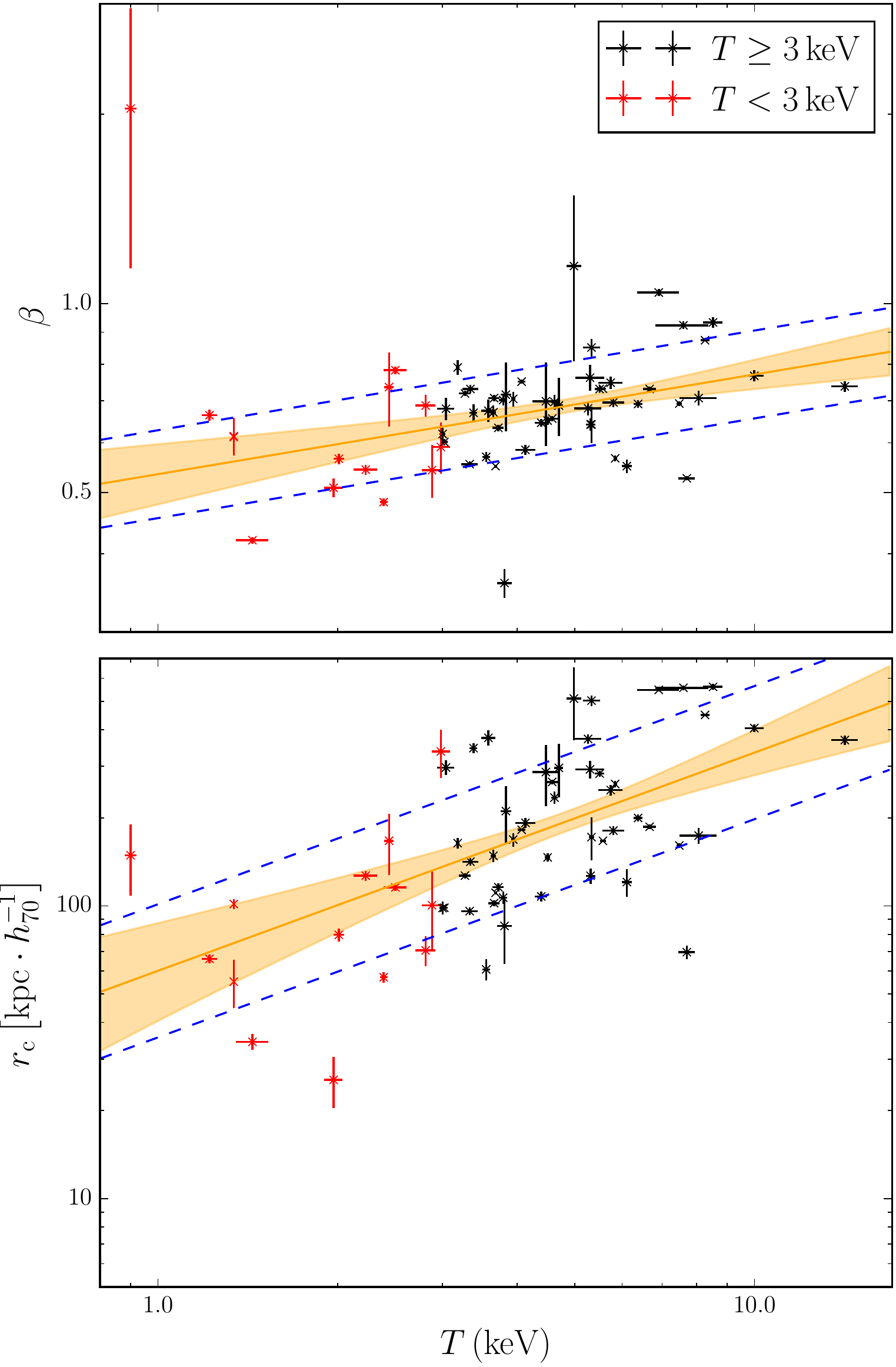}
\end{subfigure}
\caption{
Scaling relations between single $\beta$-model (\textit{left}) and core-modelled single $\beta$-model (\textit{right}) parameters and temperature. The \hiflugcs clusters with temperatures greater than \SI{3}{\keV} (\textit{black points}) are used for optimization. \textit{Red points} mark \hiflugcs objects below this temperature threshold for visualization. The orange lines and shaded regions show the best-fit relations and their uncertainties, respectively. The blue dashed lines correspond to the intrinsic log-normal scatter.}
\label{fig:scaling_relation}
\end{figure*}

\begin{figure*}
\begin{subfigure}{0.475\textwidth}
\includegraphics[width=\textwidth]{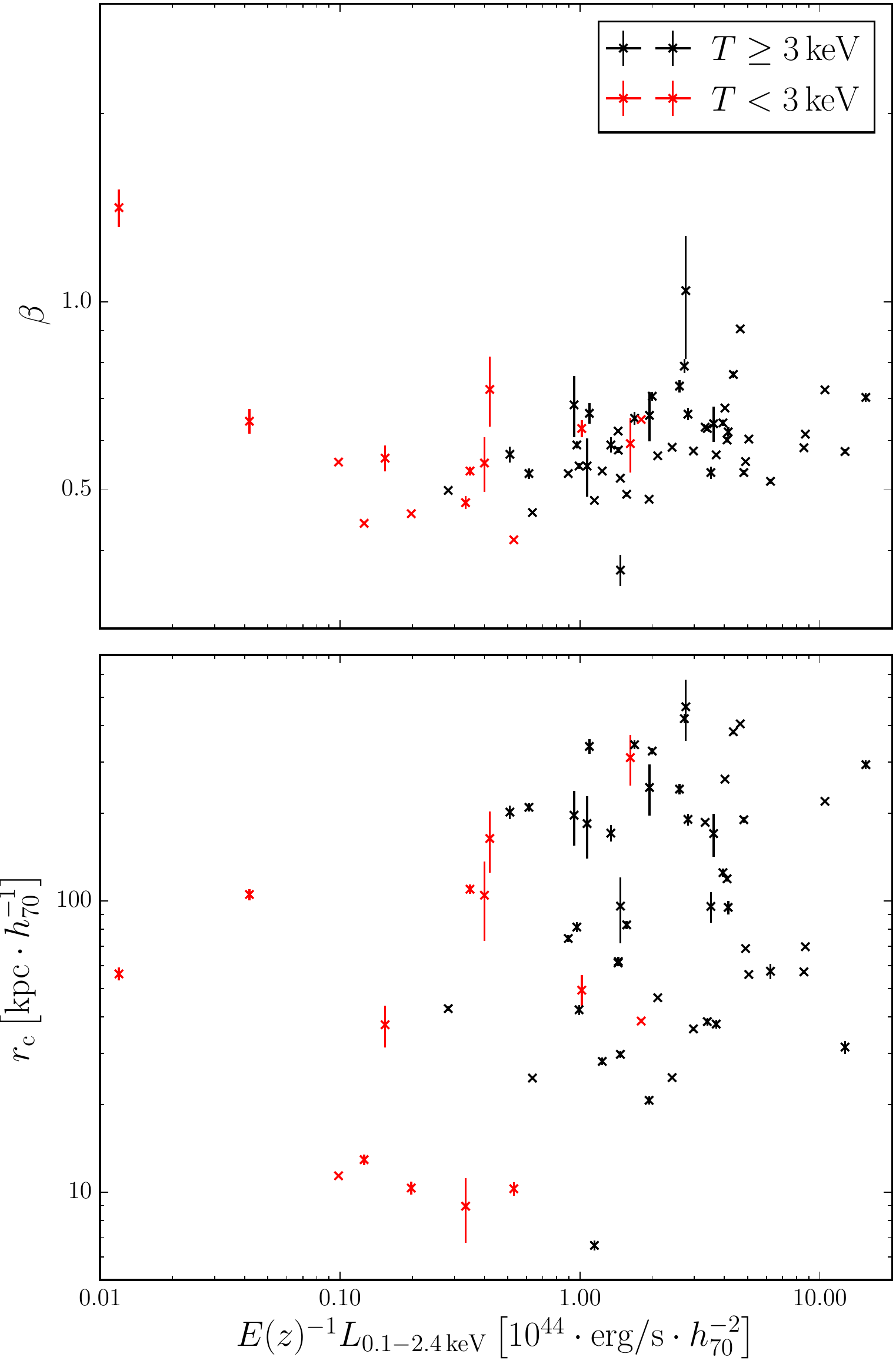}
\end{subfigure} \hspace{0.05\textwidth}
\begin{subfigure}{0.475\textwidth}
\includegraphics[width=\textwidth]{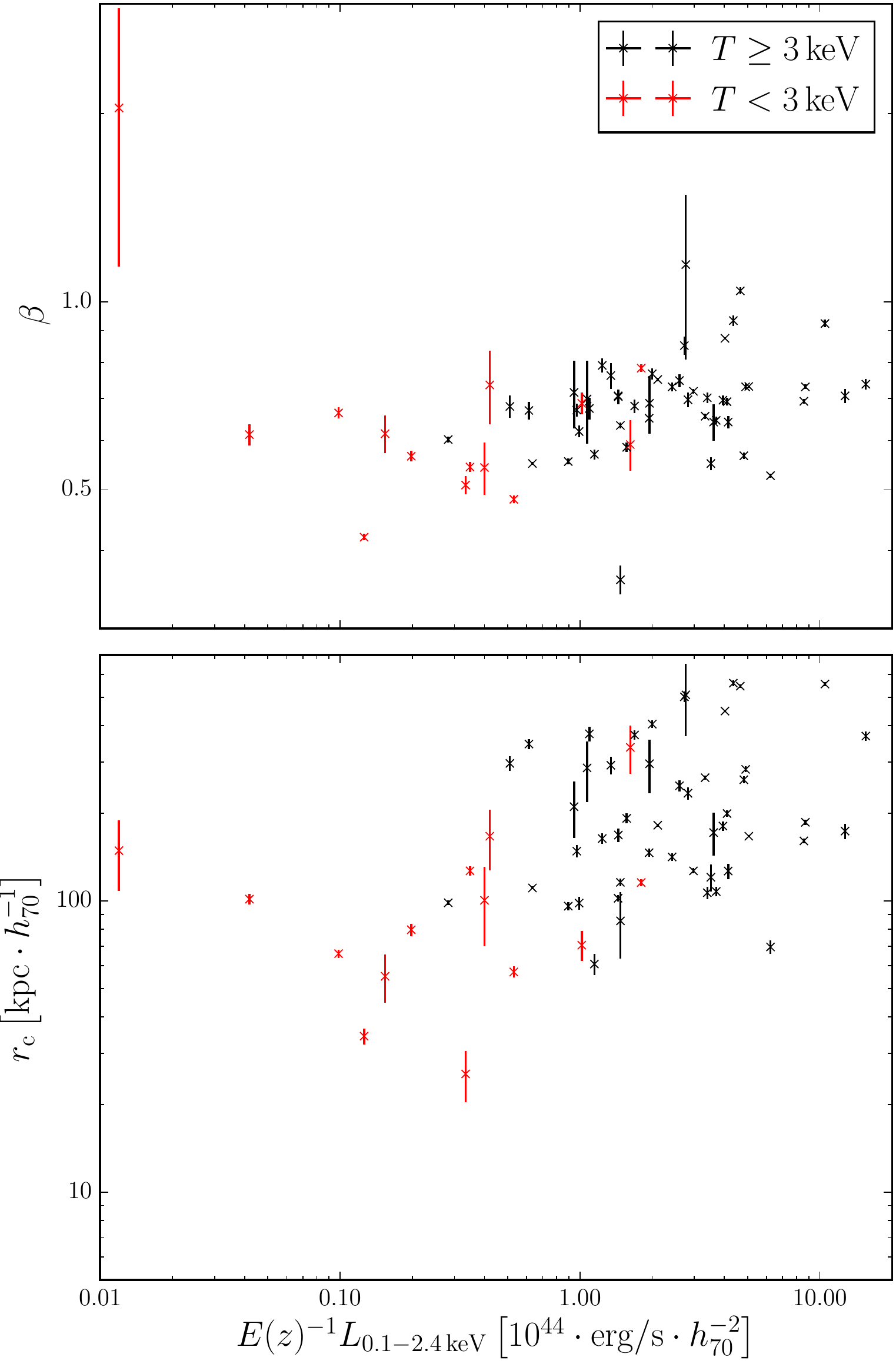}
\end{subfigure}
\caption{
Single $\beta$-model (\textit{left}) and core-modelled single $\beta$-model (\textit{right}) parameters as a function of luminosity. The \hiflugcs clusters with temperatures greater than and below \SI{3}{\keV} are marked as \textit{black} and \textit{red points}, respectively.}
\label{fig:shape_lum_relation}
\end{figure*}

This section describes how the shape parameters scale with temperature and how they correlate with luminosity. Fig. \ref{fig:rc_kT_fixed_beta_logbin} shows the core radius as a function of temperature, where the $\beta$ parameter is fixed to $2/3$ to avoid degeneracies between the surface brightness parameters. We note that the overall picture doesn't change when fixing $\beta$ to the median of the full population or the medians of the individual core-type populations. To account for a possible temperature dependence, the core radii are divided by the square-root of the corresponding cluster temperatures (see Sect. \ref{subsec:scaling_relations_discussion} for more details). The figure emphasizes the systematic differences between the individual core types in the modelling with a single $\beta$-model. The discrepancies get less prominent when modelling the core region using the wavelet decomposition. In addition, the intrinsic scatter is reduced by \numcolor{$8$\%/$11$\%/$35$\%} for NCC/WCC/SCC objects, respectively. In both modelling cases, the Spearman rank-order correlation coefficients indicate a stronger negative ($\sim -0.7$), a mild positive ($\sim 0.3$) and no significant ($\sim 0$) correlation for the NCC, WCC and SCC populations, respectively.
We determine scaling relations as outlined in Sect. \ref{subsec:scaling_relations_introduction}. The best-fit relations between shape parameters and temperature (Fig. \ref{fig:scaling_relation}), as well as luminosity and temperature are determined simultaneously. This allows studying the covariances between shape and luminosity from the joint fit. The best-fit values are shown in Table \ref{tab:scaling_relations_m_b_lambday} and \ref{tab:scaling_relations_lambdaxy}. Fig. \ref{fig:shape_lum_relation} shows the correlation between the shape parameters and luminosity. The Spearman rank-order correlation coefficients between luminosity and $\beta$/core radius are 0.37/0.12 (0.32/0.22) for the (core-modelled) single $\beta$-model. When the $\beta$ value is fixed to $2/3$, the luminosity-core radius correlation coefficients become mildly negative with -0.23 (-0.22).
\begin{table*}[htp]
\centering
\caption
{Slopes, $m$, normalizations, $b$, and intrinsic scatter values for the different $y$-temperature scaling relations for  single $\beta$-model and core-modelled single $\beta$-model parameters.
}
\label{tab:scaling_relations_m_b_lambday}
\begin{tabular}{c|ccc|ccc}
\hline
\hline
&\multicolumn{6}{c}{Single $\beta$-model} \\
\cline{2-7}
&\multicolumn{3}{c|}{} & \multicolumn{3}{c}{$\beta = 2/3$} \\
$\pmb{y}$ & $\pmb{m}$ & $\pmb{b}$ & $\pmb{\lambda_{y}}$  & $\pmb{m}$ & $\pmb{b}$ & $\pmb{\lambda_{y}}$ \\
\hline
%
%
\hline
$\beta$ & $0.19^{+0.06}_{-0.06}$ & $-0.82^{+0.09}_{-0.09}$ & $0.13^{+0.02}_{-0.01}$ & - & - & - \\
$r_{\mathrm{c}}$ & $1.04^{+0.37}_{-0.37}$ & $2.94^{+0.61}_{-0.60}$ & $0.88^{+0.10}_{-0.09}$ & $0.77^{+0.14}_{-0.14}$ & $3.73^{+0.25}_{-0.25}$ & $0.62^{+0.06}_{-0.05}$ \\
$E(z)^{-1} L_{0.1-2.4 \, \mathrm{keV}}$ & $1.90^{+0.23}_{-0.23}$ & $99.16^{+0.37}_{-0.38}$ & $0.49^{+0.07}_{-0.06}$ & $2.26^{+0.14}_{-0.14}$ & $98.58^{+0.24}_{-0.24}$ & $0.52^{+0.06}_{-0.05}$ \\
\hline
& \multicolumn{6}{c}{Core-modelled single $\beta$-model} \\
\cline{2-7}
& \multicolumn{3}{c|}{} & \multicolumn{3}{c}{$\beta = 2/3$} \\
$\pmb{y}$ & $\pmb{m}$ & $\pmb{b}$ & $\pmb{\lambda_{y}}$  & $\pmb{m}$ & $\pmb{b}$ & $\pmb{\lambda_{y}}$ \\
\hline
%
%
$\beta$ & $0.16^{+0.07}_{-0.07}$ & $-0.63^{+0.11}_{-0.11}$ & $0.16^{+0.02}_{-0.02}$ & - & - & - \\
$r_{\mathrm{c}}$ & $0.75^{+0.22}_{-0.22}$ & $4.10^{+0.36}_{-0.36}$ & $0.52^{+0.06}_{-0.05}$ & $0.50^{+0.16}_{-0.16}$ & $4.48^{+0.26}_{-0.26}$ & $0.40^{+0.05}_{-0.04}$ \\
$E(z)^{-1} L_{0.1-2.4 \, \mathrm{keV}}$ & $1.86^{+0.22}_{-0.23}$ & $99.22^{+0.37}_{-0.38}$ & $0.50^{+0.07}_{-0.06}$ & $1.90^{+0.22}_{-0.23}$ & $99.17^{+0.37}_{-0.37}$ & $0.49^{+0.07}_{-0.06}$ \\
\end{tabular}
\end{table*}
\begin{table*}[htp]
\centering
\caption
{Correlation coefficients, $\lambda_{xy}$, between the different galaxy cluster parameters $x$ and $y$ for single $\beta$-model and core-modelled single $\beta$-model parameters.
}
\label{tab:scaling_relations_lambdaxy}
\begin{tabular}{cc|cc|cc}
\hline
\hline
&& \multicolumn{2}{c|}{Single $\beta$-model} & \multicolumn{2}{c}{Core-modelled single $\beta$-model} \\
&& & $\beta = 2/3$ & & $\beta = 2/3$ \\
\cline{3-6}
$\pmb{x}$ & $\pmb{y}$ & \multicolumn{2}{c|}{$\pmb{\lambda_{xy}}$} & \multicolumn{2}{c}{$\pmb{\lambda_{xy}}$} \\
\hline
$\beta$ & $r_{\mathrm{c}}$ & $0.60^{+0.09}_{-0.11}$ & - & $0.67^{+0.08}_{-0.10}$ & - \\
$\beta$ & $E(z)^{-1} L_{0.1-2.4 \, \mathrm{keV}}$ &$-0.02^{+0.17}_{-0.17}$& - & $0.02^{+0.16}_{-0.16}$ & - \\
$r_{\mathrm{c}}$ & $E(z)^{-1} L_{0.1-2.4 \, \mathrm{keV}}$ & $-0.43^{+0.16}_{-0.13}$ & $-0.43^{+0.15}_{-0.13}$ & $-0.32^{+0.16}_{-0.15}$ & $-0.50^{+0.15}_{-0.13}$ \\
\hline
\end{tabular}
\end{table*}

\subsection{Discussion}
\label{subsec:scaling_relations_discussion}

The best-fit parameters of the single $\beta$-model and core-modelled single $\beta$-model agree on the $\sim$1$\sigma$ level, except that in the latter case the intrinsic scatter value of the core radius is reduced by almost a factor of $2$. In both modelling approaches the shape parameters show a positive correlation with temperature. The shape parameters of galaxy clusters are often fixed to generic values (\eg $\beta=2/3$) or scaling relations (\eg $r_{\textnormal{c}} \propto r_{500}$, \citet{2018A&A...620A..10P}). The assumption that the core-radius is proportional to $r_{500}$, together with equation (\ref{equ:sssTM}) results in a self-similar scaling of $r_{\textnormal{c}} \propto T^{1/2}$. We find a core-radius-temperature relation with a marginally steeper slope of \numcolor{$1.04\pm0.37$ ($0.75\pm0.20$)} for the (core-modelled) single $\beta$-model compared to this expectation. Fixing the value of $\beta$ to $2/3$ results in \numcolor{$0.77\pm0.14$ ($0.50\pm0.16$)}, consistent with the self-similar value when modelling the excess core emission. We study the correlation coefficients (Table \ref{tab:scaling_relations_lambdaxy}) between the different galaxy cluster parameters by simultaneously fitting for the scaling relations and the intrinsic scatter tensor. We expect some degeneracy between the single $\beta$-model parameters in the fitting, amongst others that a larger core-radius is compensated by a steeper slope. This is reflected in the strong positive correlation between core-radius and $\beta$. We do not find a significant correlation between $\beta$ and luminosity. There is a strong negative correlation between core radius and luminosity, \ie at a given temperature, more luminous objects tend to be more compact. For all modelling cases, the best-fit correlation coefficients are consistent on a 1$\sigma$ level. This implies, that the measured correlation between core radius and luminosity is not significantly affected by modelling the central excess emission. Neither this covariance nor the shape-temperature relations are taken into account in existing simulations for \erosita but may play a crucial role in understanding selection effects related to the detection of clusters in X-ray surveys, which is a key ingredient for the use of X-ray galaxy clusters as a precision cosmological probe. The findings presented here can be used to perform more realistic simulations and a comparison between different sets of simulations allows to study the impact of these covariances on obtained cosmological parameters.

%
\section{Summary}
\label{sec:summary}

X-ray morphologies of galaxy clusters play a crucial role in the determination of the survey selection function. We compare self-similar scaled emission measure profiles of a well defined galaxy cluster sample (\hiflugcs) above a representative temperature threshold of \SI{3}{\keV}. One outlier (Abell 3526) with small statistical errors drives the weighted mean profiles of sub-populations according to different core properties towards a different behaviour of strong cool-core and non-cool-core objects in the \SIrange[range-units = single, range-phrase=--]{0.2}{0.5}{\rfivehundred} radial range. Excluding this object from the analysis or calculating the median profiles reveals no systematic difference in the aforementioned radial range. We conclude that there is no indication for a correlation between the behaviour in the \SIrange[range-units = single, range-phrase=--]{0.2}{0.5}{\rfivehundred} radial range and the core state, although the overall shapes of the SCC and NCC populations are different. The median SCC profile shows a larger normalization towards the center and is steeper compared to the median NCC profile. This leads to a turnover of the profile ratio at $\sim$\SI{0.3}{\rfivehundred}. The difference in the center can be explained by the core state but the difference in the outskirts is still under debate. As discussed in Sect. \ref{subsec:emprofiles_discussion} possible explanations are gas redistribution between the core region and the outskirts, the current accretion of large scale blobs or different shapes of the dark matter halos leading to different concentrations at a given radius.
Characterizing galaxy cluster surface brightness profiles with single $\beta$-models is still state-of-the-art in the determination of selection functions. We investigate the residuals of a single $\beta$-model fit to the overall cluster profile, revealing that this description tends to underestimate the flux in the galaxy cluster outskirts for less extended clusters. Fixing the $\beta$ parameter to $2/3$ increases this effect dramatically (up to over $40\%$). In both cases, the core-radius measurement for SCC objects are biased low. In addition, the intrinsic scatter values with respect to the medians of the self-similar scaled extent parameters show a more than $1\sigma$ tension between strong and non-cool-core objects.
These three effects can be minimized by adapting a wavelet decomposition based surface brightness modelling that is sensitive to the galaxy cluster outskirts and models the excess emission in the core region. Then, the fit is not driven by the local processes in the core. Compared to a single $\beta$-model approach the residuals in the \SIrange[range-units = single, range-phrase=--]{0.2}{0.5}{\rfivehundred} radial range are much smaller and the core radii depend much more mildly on the core state. Our method to model the excess core emission has very interesting applications for future galaxy cluster surveys, \eg with \erosita. The performance study for high redshift objects with small angular extent and establishing the most robust method for clusters where $r_{500}$ is unknown is left to a future work.
Using wavelet decomposition allows to determine large scale ellipticities of the clusters. The ellipticity is an interesting new survey measure for \erosita since it's determination doesn't require many photon counts and it adds additional information to the $\beta$-model shape parameters and core-excised luminosities. A detailed study of measuring galaxy cluster ellipticities with \erosita and it's implications is left to a future work.
We study how shape parameters and luminosity scale with temperature. There is no significant difference of the best-fit values between a single $\beta$-model and core-modelled single $\beta$-model, except that the intrinsic scatter of the core radius is almost twice as large for the single $\beta$-model case. The slope of the  core radius-temperature relation is steeper than the self-similar prediction of $1/2$ but gets in agreement when fixing the $\beta$ parameter to $2/3$ in the surface brightness profile modelling. More interestingly, the shape parameters are covarient with luminosity, \ie at a given temperature, more compact objects are more luminous. These covariances are usually neglected in simulations to determine the survey selection function \citep{2007MNRAS.382.1289P,2018A&A...617A..92C}. In addition, these previous studies assumed a fixed $\beta$ value, while we find that $\beta$ is a function of temperature. Taking shape-temperature scaling relations and shape-luminosity covariances into account will lead to a more realistic set of simulated galaxy clusters and will provide a better understanding of the survey completeness.

%
\section*{Acknowledgements}
This research was made possible by the International Max Planck Research School on Astrophysics at the Ludwig-Maximilians University Munich, as well as their funding received from the Max-Planck Society.

%
\bibliographystyle{aa}
\bibliography{AA_2019_35124}

\begin{thebibliography}{56}
\expandafter\ifx\csname natexlab\endcsname\relax\def\natexlab#1{#1}\fi

\bibitem[{{Allgood} {et~al.}(2006){Allgood}, {Flores}, {Primack}, {Kravtsov},
  {Wechsler}, {Faltenbacher}, \& {Bullock}}]{2006MNRAS.367.1781A}
{Allgood}, B., {Flores}, R.~A., {Primack}, J.~R., {et~al.} 2006, Monthly
  Notices of the RAS, 367, 1781

\bibitem[{{Anders} \& {Grevesse}(1989)}]{1989GeCoA..53..197A}
{Anders}, E. \& {Grevesse}, N. 1989, Geochimica Cosmochimica Acta, 53, 197

\bibitem[{{Arnaud} {et~al.}(2002){Arnaud}, {Aghanim}, \&
  {Neumann}}]{2002A&A...389....1A}
{Arnaud}, M., {Aghanim}, N., \& {Neumann}, D.~M. 2002, Astronomy and
  Astrophysics, 389, 1

\bibitem[{{Bartelmann} \& {Steinmetz}(1996)}]{1996MNRAS.283..431B}
{Bartelmann}, M. \& {Steinmetz}, M. 1996, Monthly Notices of the RAS, 283, 431

\bibitem[{{Bertin} \& {Arnouts}(1996)}]{1996A&AS..117..393B}
{Bertin}, E. \& {Arnouts}, S. 1996, Astronomy and Astrophysics, Supplement,
  117, 393

\bibitem[{{Bianconi} {et~al.}(2013){Bianconi}, {Ettori}, \&
  {Nipoti}}]{2013MNRAS.434.1565B}
{Bianconi}, M., {Ettori}, S., \& {Nipoti}, C. 2013, Monthly Notices of the RAS,
  434, 1565

\bibitem[{{Boese}(2000)}]{2000A&AS..141..507B}
{Boese}, F.~G. 2000, Astronomy and Astrophysics, Supplement, 141, 507

\bibitem[{{B{\"o}hringer} {et~al.}(2000){B{\"o}hringer}, {Voges}, {Huchra},
  {McLean}, {Giacconi}, {Rosati}, {Burg}, {Mader}, {Schuecker}, {Simi{\c c}},
  {Komossa}, {Reiprich}, {Retzlaff}, \& {Tr{\"u}mper}}]{2000ApJS..129..435B}
{B{\"o}hringer}, H., {Voges}, W., {Huchra}, J.~P., {et~al.} 2000, The
  Astrophysical Journal, Supplement, 129, 435

\bibitem[{{Bryan} \& {Norman}(1998)}]{1998ApJ...495...80B}
{Bryan}, G.~L. \& {Norman}, M.~L. 1998, Astrophysical Journal, 495, 80

\bibitem[{{Cash}(1979)}]{1979ApJ...228..939C}
{Cash}, W. 1979, Astrophysical Journal, 228, 939

\bibitem[{{Cavaliere} \& {Fusco-Femiano}(1976)}]{1976A&A....49..137C}
{Cavaliere}, A. \& {Fusco-Femiano}, R. 1976, Astronomy and Astrophysics, 49,
  137

\bibitem[{{Chen} {et~al.}(2007){Chen}, {Reiprich}, {B{\"o}hringer}, {Ikebe}, \&
  {Zhang}}]{2007A&A...466..805C}
{Chen}, Y., {Reiprich}, T.~H., {B{\"o}hringer}, H., {Ikebe}, Y., \& {Zhang},
  Y.-Y. 2007, Astronomy and Astrophysics, 466, 805

\bibitem[{{Clerc} {et~al.}(2018){Clerc}, {Ramos-Ceja}, {Ridl}, {Lamer},
  {Brunner}, {Hofmann}, {Comparat}, {Pacaud}, {K{\"a}fer}, {Reiprich},
  {Merloni}, {Schmid}, {Brand}, {Wilms}, {Friedrich}, {Finoguenov}, {Dauser},
  \& {Kreykenbohm}}]{2018A&A...617A..92C}
{Clerc}, N., {Ramos-Ceja}, M.~E., {Ridl}, J., {et~al.} 2018, Astronomy and
  Astrophysics, 617, A92

\bibitem[{{Eckert} {et~al.}(2011){Eckert}, {Molendi}, \&
  {Paltani}}]{2011A&A...526A..79E}
{Eckert}, D., {Molendi}, S., \& {Paltani}, S. 2011, Astronomy and Astrophysics,
  526, A79

\bibitem[{{Eckert} {et~al.}(2012){Eckert}, {Vazza}, {Ettori}, {Molendi},
  {Nagai}, {Lau}, {Roncarelli}, {Rossetti}, {Snowden}, \&
  {Gastaldello}}]{2012A&A...541A..57E}
{Eckert}, D., {Vazza}, F., {Ettori}, S., {et~al.} 2012, Astronomy and
  Astrophysics, 541, A57

\bibitem[{{Foreman-Mackey} {et~al.}(2013){Foreman-Mackey}, {Hogg}, {Lang}, \&
  {Goodman}}]{2013PASP..125..306F}
{Foreman-Mackey}, D., {Hogg}, D.~W., {Lang}, D., \& {Goodman}, J. 2013,
  Publications of the ASP, 125, 306

\bibitem[{{Ghirardini} {et~al.}(2018){Ghirardini}, {Eckert}, {Ettori},
  {Pointecouteau}, {Molendi}, {Gaspari}, {Rossetti}, {De Grandi}, {Roncarelli},
  {Bourdin}, {Mazzotta}, {Rasia}, \& {Vazza}}]{2018arXiv180500042G}
{Ghirardini}, V., {Eckert}, D., {Ettori}, S., {et~al.} 2018, ArXiv e-prints
  [\eprint[arXiv]{1805.00042}]

\bibitem[{{Giodini} {et~al.}(2013){Giodini}, {Lovisari}, {Pointecouteau},
  {Ettori}, {Reiprich}, \& {Hoekstra}}]{2013SSRv..177..247G}
{Giodini}, S., {Lovisari}, L., {Pointecouteau}, E., {et~al.} 2013, Space
  Science Reviews, 177, 247

\bibitem[{{Greiner} {et~al.}(2016){Greiner}, {Burgess}, {Savchenko}, \&
  {Yu}}]{2016ApJ...827L..38G}
{Greiner}, J., {Burgess}, J.~M., {Savchenko}, V., \& {Yu}, H.-F. 2016,
  Astrophysical Journal, Letters, 827, L38

\bibitem[{{Hashimoto} {et~al.}(2007){Hashimoto}, {B{\"o}hringer}, {Henry},
  {Hasinger}, \& {Szokoly}}]{2007A&A...467..485H}
{Hashimoto}, Y., {B{\"o}hringer}, H., {Henry}, J.~P., {Hasinger}, G., \&
  {Szokoly}, G. 2007, Astronomy and Astrophysics, 467, 485

\bibitem[{{Hudson} {et~al.}(2010){Hudson}, {Mittal}, {Reiprich}, {Nulsen},
  {Andernach}, \& {Sarazin}}]{2010A&A...513A..37H}
{Hudson}, D.~S., {Mittal}, R., {Reiprich}, T.~H., {et~al.} 2010, Astronomy and
  Astrophysics, 513, A37

\bibitem[{{Kaiser}(1986)}]{1986MNRAS.222..323K}
{Kaiser}, N. 1986, Monthly Notices of the RAS, 222, 323

\bibitem[{{Kelly}(2007)}]{2007ApJ...665.1489K}
{Kelly}, B.~C. 2007, Astrophysical Journal, 665, 1489

\bibitem[{{King}(1962)}]{1962AJ.....67..471K}
{King}, I. 1962, The Astronomical Journal, 67, 471

\bibitem[{{Kravtsov} {et~al.}(2006){Kravtsov}, {Vikhlinin}, \&
  {Nagai}}]{2006ApJ...650..128K}
{Kravtsov}, A.~V., {Vikhlinin}, A., \& {Nagai}, D. 2006, Astrophysical Journal,
  650, 128

\bibitem[{{Lau} {et~al.}(2012){Lau}, {Nagai}, {Kravtsov}, {Vikhlinin}, \&
  {Zentner}}]{2012ApJ...755..116L}
{Lau}, E.~T., {Nagai}, D., {Kravtsov}, A.~V., {Vikhlinin}, A., \& {Zentner},
  A.~R. 2012, Astrophysical Journal, 755, 116

\bibitem[{{Lin} {et~al.}(2015){Lin}, {McDonald}, {Benson}, \&
  {Miller}}]{2015ApJ...802...34L}
{Lin}, H.~W., {McDonald}, M., {Benson}, B., \& {Miller}, E. 2015, Astrophysical
  Journal, 802, 34

\bibitem[{{Macci{\`o}} {et~al.}(2008){Macci{\`o}}, {Dutton}, \& {van den
  Bosch}}]{2008MNRAS.391.1940M}
{Macci{\`o}}, A.~V., {Dutton}, A.~A., \& {van den Bosch}, F.~C. 2008, Monthly
  Notices of the RAS, 391, 1940

\bibitem[{{Mantz} {et~al.}(2010){Mantz}, {Allen}, {Ebeling}, {Rapetti}, \&
  {Drlica-Wagner}}]{2010MNRAS.406.1773M}
{Mantz}, A., {Allen}, S.~W., {Ebeling}, H., {Rapetti}, D., \& {Drlica-Wagner},
  A. 2010, Monthly Notices of the RAS, 406, 1773

\bibitem[{{Mantz} {et~al.}(2016){Mantz}, {Allen}, {Morris}, {von der Linden},
  {Applegate}, {Kelly}, {Burke}, {Donovan}, \& {Ebeling}}]{2016MNRAS.463.3582M}
{Mantz}, A.~B., {Allen}, S.~W., {Morris}, R.~G., {et~al.} 2016, Monthly Notices
  of the RAS, 463, 3582

\bibitem[{{Maughan}(2007)}]{2007ApJ...668..772M}
{Maughan}, B.~J. 2007, Astrophysical Journal, 668, 772

\bibitem[{{Maughan} {et~al.}(2012){Maughan}, {Giles}, {Randall}, {Jones}, \&
  {Forman}}]{2012MNRAS.421.1583M}
{Maughan}, B.~J., {Giles}, P.~A., {Randall}, S.~W., {Jones}, C., \& {Forman},
  W.~R. 2012, Monthly Notices of the RAS, 421, 1583

\bibitem[{{Merloni} {et~al.}(2012){Merloni}, {Predehl}, {Becker},
  {B{\"o}hringer}, {Boller}, {Brunner}, {Brusa}, {Dennerl}, {Freyberg},
  {Friedrich}, {Georgakakis}, {Haberl}, {Hasinger}, {Meidinger}, {Mohr},
  {Nandra}, {Rau}, {Reiprich}, {Robrade}, {Salvato}, {Santangelo}, {Sasaki},
  {Schwope}, {Wilms}, \& {German eROSITA Consortium}}]{2012arXiv1209.3114M}
{Merloni}, A., {Predehl}, P., {Becker}, W., {et~al.} 2012, ArXiv e-prints
  [\eprint[arXiv]{1209.3114}]

\bibitem[{{Mittal} {et~al.}(2011){Mittal}, {Hicks}, {Reiprich}, \&
  {Jaritz}}]{2011A&A...532A.133M}
{Mittal}, R., {Hicks}, A., {Reiprich}, T.~H., \& {Jaritz}, V. 2011, Astronomy
  and Astrophysics, 532, A133

\bibitem[{{Navarro} {et~al.}(1995){Navarro}, {Frenk}, \&
  {White}}]{1995MNRAS.275..720N}
{Navarro}, J.~F., {Frenk}, C.~S., \& {White}, S.~D.~M. 1995, Monthly Notices of
  the RAS, 275, 720

\bibitem[{{Navarro} {et~al.}(1996){Navarro}, {Frenk}, \&
  {White}}]{1996ApJ...462..563N}
{Navarro}, J.~F., {Frenk}, C.~S., \& {White}, S.~D.~M. 1996, Astronomy and
  Astrophysics, 462, 563

\bibitem[{{Navarro} {et~al.}(1997){Navarro}, {Frenk}, \&
  {White}}]{1997ApJ...490..493N}
{Navarro}, J.~F., {Frenk}, C.~S., \& {White}, S.~D.~M. 1997, Astronomy and
  Astrophysics, 490, 493

\bibitem[{{Nelson} {et~al.}(2014){Nelson}, {Lau}, \&
  {Nagai}}]{2014ApJ...792...25N}
{Nelson}, K., {Lau}, E.~T., \& {Nagai}, D. 2014, Astrophysical Journal, 792, 25

\bibitem[{{Pacaud} {et~al.}(2007){Pacaud}, {Pierre}, {Adami}, {Altieri},
  {Andreon}, {Chiappetti}, {Detal}, {Duc}, {Galaz}, {Gueguen}, {Le F{\`e}vre},
  {Hertling}, {Libbrecht}, {Melin}, {Ponman}, {Quintana}, {Refregier},
  {Sprimont}, {Surdej}, {Valtchanov}, {Willis}, {Alloin}, {Birkinshaw},
  {Bremer}, {Garcet}, {Jean}, {Jones}, {Le F{\`e}vre}, {Maccagni}, {Mazure},
  {Proust}, {R{\"o}ttgering}, \& {Trinchieri}}]{2007MNRAS.382.1289P}
{Pacaud}, F., {Pierre}, M., {Adami}, C., {et~al.} 2007, Monthly Notices of the
  RAS, 382, 1289

\bibitem[{{Pacaud} {et~al.}(2018){Pacaud}, {Pierre}, {Melin}, {Adami},
  {Evrard}, {Galli}, {Gastaldello}, {Maughan}, {Sereno}, {Alis}, {Altieri},
  {Birkinshaw}, {Chiappetti}, {Faccioli}, {Giles}, {Horellou}, {Iovino},
  {Koulouridis}, {Le F{\`e}vre}, {Lidman}, {Lieu}, {Maurogordato},
  {Moscardini}, {Poggianti}, {Pompei}, {Sadibekova}, {Valtchanov}, \&
  {Willis}}]{2018A&A...620A..10P}
{Pacaud}, F., {Pierre}, M., {Melin}, J.-B., {et~al.} 2018, Astronomy and
  Astrophysics, 620, A10

\bibitem[{{Planck Collaboration} {et~al.}(2016){Planck Collaboration}, {Ade},
  {Aghanim}, {Arnaud}, {Ashdown}, {Aumont}, {Baccigalupi}, {Banday},
  {Barreiro}, {Barrena}, \& et~al.}]{2016A&A...594A..27P}
{Planck Collaboration}, {Ade}, P.~A.~R., {Aghanim}, N., {et~al.} 2016,
  Astronomy and Astrophysics, 594, A27

\bibitem[{{Pratt} {et~al.}(2009){Pratt}, {Croston}, {Arnaud}, \&
  {B{\"o}hringer}}]{2009A&A...498..361P}
{Pratt}, G.~W., {Croston}, J.~H., {Arnaud}, M., \& {B{\"o}hringer}, H. 2009,
  Astronomy and Astrophysics, 498, 361

\bibitem[{{Predehl} {et~al.}(2018){Predehl}, {Bornemann}, {Br{\"a}uninger},
  {Brunner}, {Burwitz}, {Coutinho}, {Dennerl}, {Eder}, {Friedrich},
  {F{\"u}rmetz}, {Hartner}, {von Kienlin}, {Kink}, {Meidinger}, {Mican},
  {M{\"u}ller}, {Nandra}, {Pfeffermann}, {Roh{\'e}}, \&
  {Yaroshenko}}]{2018SPIE10699E..5HP}
{Predehl}, P., {Bornemann}, W., {Br{\"a}uninger}, H., {et~al.} 2018, in Society
  of Photo-Optical Instrumentation Engineers (SPIE) Conference Series, Vol.
  10699, Space Telescopes and Instrumentation 2018: Ultraviolet to Gamma Ray,
  106995H

\bibitem[{{Reiprich} \& {B{\"o}hringer}(2002)}]{2002ApJ...567..716R}
{Reiprich}, T.~H. \& {B{\"o}hringer}, H. 2002, Astrophysical Journal, 567, 716

\bibitem[{{Rossetti} {et~al.}(2017){Rossetti}, {Gastaldello}, {Eckert}, {Della
  Torre}, {Pantiri}, {Cazzoletti}, \& {Molendi}}]{2017MNRAS.468.1917R}
{Rossetti}, M., {Gastaldello}, F., {Eckert}, D., {et~al.} 2017, Monthly Notices
  of the RAS, 468, 1917

\bibitem[{{Sanderson} {et~al.}(2006){Sanderson}, {Ponman}, \&
  {O'Sullivan}}]{2006MNRAS.372.1496S}
{Sanderson}, A.~J.~R., {Ponman}, T.~J., \& {O'Sullivan}, E. 2006, Monthly
  Notices of the RAS, 372, 1496

\bibitem[{{Santos} {et~al.}(2008){Santos}, {Rosati}, {Tozzi}, {B{\"o}hringer},
  {Ettori}, \& {Bignamini}}]{2008A&A...483...35S}
{Santos}, J.~S., {Rosati}, P., {Tozzi}, P., {et~al.} 2008, Astronomy and
  Astrophysics, 483, 35

\bibitem[{{Schellenberger} \& {Reiprich}(2017)}]{2017MNRAS.469.3738S}
{Schellenberger}, G. \& {Reiprich}, T.~H. 2017, Monthly Notices of the RAS,
  469, 3738

\bibitem[{{Snowden} {et~al.}(1994){Snowden}, {McCammon}, {Burrows}, \&
  {Mendenhall}}]{1994ApJ...424..714S}
{Snowden}, S.~L., {McCammon}, D., {Burrows}, D.~N., \& {Mendenhall}, J.~A.
  1994, Astrophysical Journal, 424, 714

\bibitem[{VanderPlas(2016)}]{VanderPlas:2016:PDS:3133359}
VanderPlas, J. 2016, Python Data Science Handbook: Essential Tools for Working
  with Data, 1st edn. (O'Reilly Media, Inc.)

\bibitem[{{Vikhlinin} {et~al.}(2007){Vikhlinin}, {Burenin}, {Forman}, {Jones},
  {Hornstrup}, {Murray}, \& {Quintana}}]{2007hvcg.conf...48V}
{Vikhlinin}, A., {Burenin}, R., {Forman}, W.~R., {et~al.} 2007, in Heating
  versus Cooling in Galaxies and Clusters of Galaxies, ed. H.~{B{\"o}hringer},
  G.~W. {Pratt}, A.~{Finoguenov}, \& P.~{Schuecker}, 48

\bibitem[{{Vikhlinin} {et~al.}(1999){Vikhlinin}, {Forman}, \&
  {Jones}}]{1999ApJ...525...47V}
{Vikhlinin}, A., {Forman}, W., \& {Jones}, C. 1999, Astrophysical Journal, 525,
  47

\bibitem[{{Vikhlinin} {et~al.}(1998){Vikhlinin}, {McNamara}, {Forman}, {Jones},
  {Quintana}, \& {Hornstrup}}]{1998ApJ...502..558V}
{Vikhlinin}, A., {McNamara}, B.~R., {Forman}, W., {et~al.} 1998, Astrophysical
  Journal, 502, 558

\bibitem[{{Voges} {et~al.}(1999){Voges}, {Aschenbach}, {Boller},
  {Br{\"a}uninger}, {Briel}, {Burkert}, {Dennerl}, {Englhauser}, {Gruber},
  {Haberl}, {Hartner}, {Hasinger}, {K{\"u}rster}, {Pfeffermann}, {Pietsch},
  {Predehl}, {Rosso}, {Schmitt}, {Tr{\"u}mper}, \&
  {Zimmermann}}]{1999A&A...349..389V}
{Voges}, W., {Aschenbach}, B., {Boller}, T., {et~al.} 1999, Astronomy and
  Astrophysics, 349, 389

\bibitem[{{Willingale} {et~al.}(2013){Willingale}, {Starling}, {Beardmore},
  {Tanvir}, \& {O'Brien}}]{2013MNRAS.431..394W}
{Willingale}, R., {Starling}, R.~L.~C., {Beardmore}, A.~P., {Tanvir}, N.~R., \&
  {O'Brien}, P.~T. 2013, Monthly Notices of the RAS, 431, 394

\bibitem[{{Zhang} {et~al.}(2007){Zhang}, {Finoguenov}, {B{\"o}hringer},
  {Kneib}, {Smith}, {Czoske}, \& {Soucail}}]{2007A&A...467..437Z}
{Zhang}, Y.-Y., {Finoguenov}, A., {B{\"o}hringer}, H., {et~al.} 2007, Astronomy
  and Astrophysics, 467, 437

\end{thebibliography}

%
\begin{appendix}

%
\section{Mass comparison}

\begin{figure}
\centering
\resizebox{\hsize}{!}{\includegraphics{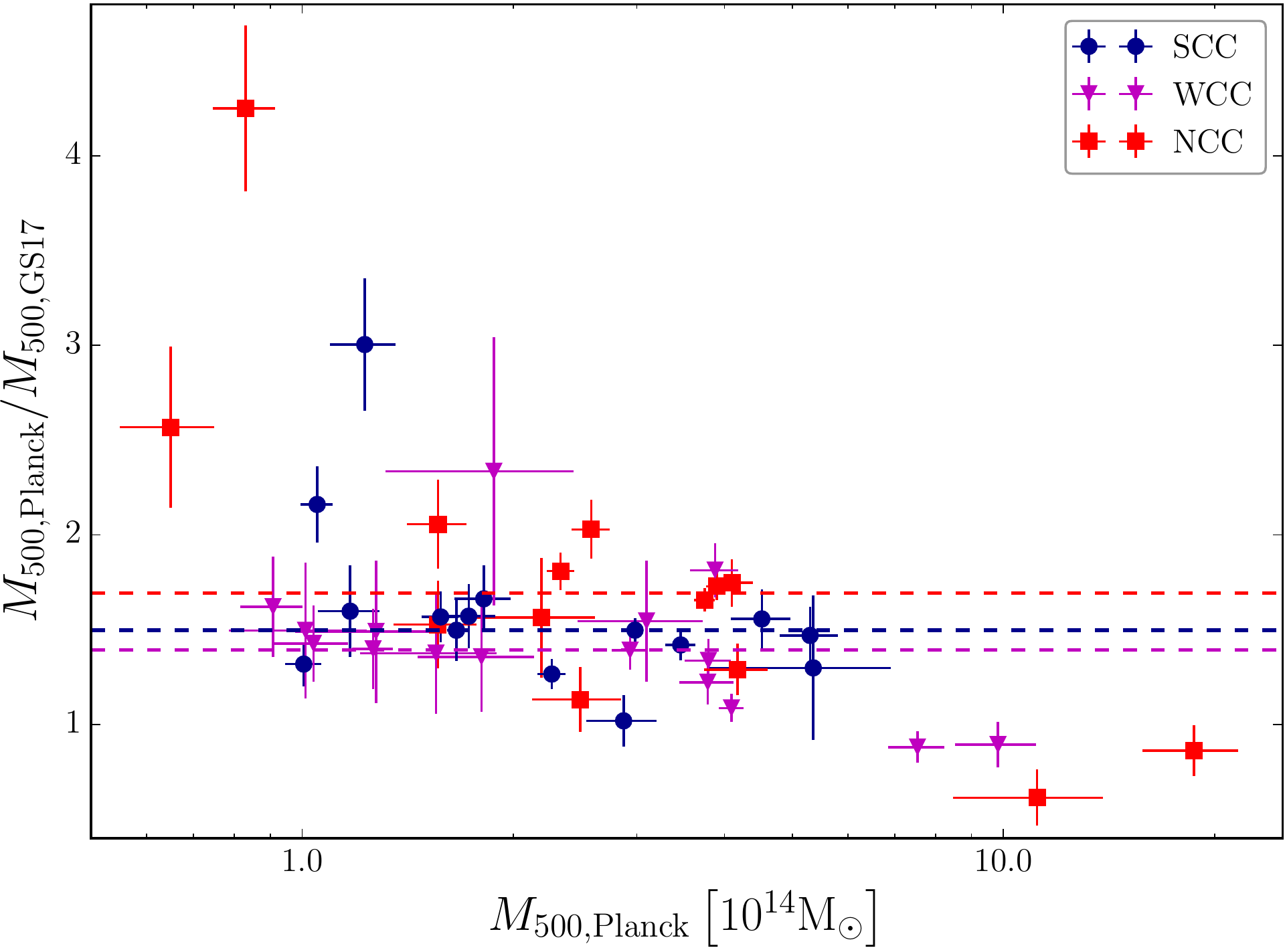}}
\caption{Comparison between Planck and hydrostatic mass estimates of \hiflugcs objects with temperatures greater than \SI{3}{\keV}.
}
\label{fig:m500_comparisson_Planck_GS17}
\end{figure}

In this section we compare our previous emission measure ratio results of Sect. \ref{sec:emprofiles} to profiles that are re-scaled by a characteristic radii according to hydrostatic mass estimates by \citet{2017MNRAS.469.3738S}. We adapt their preferred "NFW Freeze" model, where a NFW profile \citep{1996ApJ...462..563N,1997ApJ...490..493N} is fit to the outermost measured mass profiles of Chandra observations and a concentration-mass relation is used to reduce the degrees of freedom. We are interested in the difference between individual core-types and not in the bias between the Planck and hydrostatic masses. Therefore we assume that the bias is constant for all clusters. To probe the masses at the same radii, we recalculate the hydrostatic masses at the Planck $r_{500}$ values according to Formula (31) of \citet{2017MNRAS.469.3738S}
\begin{equation}
M^{*}_{500,\textnormal{GS17}}\left(<r_{500,\textnormal{Planck}}\right) = M_{500,\textnormal{GS17}}
\frac{Y(\frac{r_{500,\textnormal{Planck}}}{r_{500,\textnormal{GS17}}}c_{500,\textnormal{GS17}})}{Y(c_{500,\textnormal{GS17}})},
\label{equ:Mcomparison}
\end{equation}
where $c_{500}$ denotes the NFW concentration parameter and $Y(u)=\ln(1+u)-u/(1+u)$. These recalculated masses are on average lower than the corresponding masses in the Planck catalogue. The median of the NCC cluster masses is \numcolor{$\sim$13\%} larger than that of the SCC objects (see Fig. \ref{fig:m500_comparisson_Planck_GS17}), increasing the SCC to NCC weighted mean ratio effect in the \SIrange[range-units = single, range-phrase=--]{0.2}{0.5}{\rfivehundred} range to $40$\%. This increase due to the dynamical states is also seen in simulations, \eg in different fractions of non-thermal pressure for different mass accretion rates \citep{2014ApJ...792...25N}. The differences can thus be explained by assuming that NCC clusters are merging objects and that they contain more non-thermal energy.
An alternative explanation is that the differences come from the assumed concentration-mass relation. We expect different core-types to have different shapes of the dark matter halo and therefore different concentrations at a given radius. However, fitting for the concentration in \citet{2017MNRAS.469.3738S} results in some cases to unrealistic high or low masses, potentially because of limited radial coverage due to the relatively small \chandra field-of-view. 

%
\section{Priors}
\label{sec:priors}

Table \ref{tab:param_priors} shows the priors used for this analysis. They are chosen to be weakly- or non-informative and varying them does not influence the results of this study significantly.
\begin{table}[htp]
\centering
\renewcommand*{\arraystretch}{2.5}
\caption
{List of parameters and their priors. We note that the prior of the scaling relation slope is assumed to be uniform in $\sin(\Theta)$ \citep{VanderPlas:2016:PDS:3133359}, with $\Theta$ being the angle between the best-fit line and the $x$-axis. The term "pos-normal" refers to a probability distribution that follows an ordinary normal distribution but is set to zero for negative parameter values, \ie the parameter is restricted to be positive.
}
\label{tab:param_priors}
\begin{tabular}{ccc}
\hline
\hline
\textbf{Parameter} & \textbf{Description} & \textbf{Prior} \\
\hline
$\beta$ & $\beta$-model slope & pos-normal(0.67,10) \\
$r_{\textnormal{c}}$ & \makecell{$\beta$-model core radius \\ in arcminutes} & pos-normal(0,100) \\
$m$ & Scaling relation slope & $\propto \left(1+m^{2}\right)^{-3/2}$\\
$b$ & Scaling relation intercept & \makecell{non-informative \\ uniform}  \\
$\lambda_{y}$ & \makecell{Intrinsic scatter \\ in $y$} & pos-normal(0,10) \\
$\lambda_{xy}$ & \makecell{Correlation coefficient \\ between $x$ and $y$} & uniform[-1,1] \\
\hline
\end{tabular}
\end{table}
%

%
\section{\hiflugcs parameters}
\onecolumn
\begin{landscape}
\begin{longtable}{ccccccccccccc}
\caption{Galaxy cluster parameters. Column (1)
gives the cluster name. Columns (2)
and (3) list the equatorial coordinates of the cluster center in decimal degrees based on the large scale wavelet image. Column (4) gives the offset to an iteratively determined two-dimensional 'center of mass' using an aperture radius of \SI{3}{\arcmin} \citep{2002ApJ...567..716R}. Column (5) and (6) list the cluster redshift \citep{2002ApJ...567..716R} and core-excised temperature \citep{2010A&A...513A..37H}, respectively. 
Column (7) and (8) give the $\beta$-model slope and core radius for a core-modelled fit. Column (9) lists the luminosity in the \SIrange[range-units = single, range-phrase=--]{0.1}{2.4}{\keV} energy range. The cool-core classification according to \citet{2010A&A...513A..37H} is given in column (10). Column (11) lists the characteristic radius where the density corresponds to 500 times the critical density. Galaxy clusters, whose $r_{500}$ value is marked with a $^{\dagger}$ don't have SZ mass estimates \citep{2016A&A...594A..27P} and the \citet{2017MNRAS.469.3738S} mass estimate is used to determine the wavelet small scales. These clusters are excluded from further analysis steps which comprise characteristic radii. Column (12) gives the measured large scale ellipticity. The physical to angular scale conversion at the cluster redshift is given in column (13).
}\\
\hline\hline
\textbf{Cluster} & \textbf{R.A.} & \textbf{Dec.} & \textbf{Offset} & $\pmb{z}$ & $\pmb{T}$ & $\pmb{\beta}$ & $\pmb{r_{\textnormal{\textbf{c}}}}$ & $\pmb{L}$ & \textbf{CCC} & $\pmb{r_{500}}$ & \textbf{Ellipticity} & \textbf{Scale} \\
\textbf{name} & \multicolumn{2}{c}{J2000} & \SI{}{\arcsec} & & \SI{}{\keV} & & \SI{}{\kilo\parsec} & \SI{E44}{\erg\per\second} & & \SI{}{\mega\parsec} & & \SI{}{\kilo\parsec\per\arcsec} \\
\hline
\endfirsthead
\caption{Continued.}\\
\hline\hline
\textbf{Cluster} & \textbf{R.A.} & \textbf{Dec.} & \textbf{Offset} & $\pmb{z}$ & $\pmb{T}$ & $\pmb{\beta}$ & $\pmb{r_{\textnormal{\textbf{c}}}}$ & $\pmb{L}$ & \textbf{CCC} & $\pmb{r_{500}}$ & \textbf{Ellipticity} & \textbf{Scale} \\
\textbf{name} & \multicolumn{2}{c}{J2000} & \SI{}{\arcsec} & & \SI{}{\keV} & & \SI{}{\kilo\parsec} & \SI{E44}{\erg\per\second} & & \SI{}{\mega\parsec} &  & \SI{}{\kilo\parsec\per\arcsec} \\
\hline
\endhead
\hline
A0085 & 10.4600 & -9.3110 & 23.5 & 0.0556 & $5.50_{-0.10}^{+0.10}$ & $0.731_{-0.009}^{+0.010}$ & $283.33_{-6.43}^{+6.72}$ & $5.034\pm0.030$ & SCC & 1.18 & 0.102 & 1.08 \\
A0119 & 14.0610 & -1.2490 & 14.1 & 0.0440 & $5.26_{-0.26}^{+0.30}$ & $0.681_{-0.017}^{+0.017}$ & $371.94_{-13.22}^{+13.49}$ & $1.725\pm0.016$ & NCC & 1.05 & 0.177 & 0.87 \\
A0133 & 15.6740 & -21.8800 & 1.9 & 0.0569 & $3.72_{-0.09}^{+0.07}$ & $0.634_{-0.007}^{+0.008}$ & $115.90_{-3.59}^{+3.79}$ & $1.514\pm0.012$ & SCC & 1.00 & 0.240 & 1.10 \\
NGC507 & 20.9080 & 33.2560 & 9.1 & 0.0165 & $1.44_{-0.10}^{+0.08}$ & $0.420_{-0.005}^{+0.005}$ & $34.35_{-2.10}^{+2.17}$ & $0.127\pm0.002$ & SCC & $0.47^{\dagger}$ & 0.047 & 0.34 \\
A0262 & 28.1960 & 36.1600 & 26.2 & 0.0161 & $2.39_{-0.04}^{+0.03}$ & $0.483_{-0.006}^{+0.007}$ & $57.07_{-2.45}^{+2.55}$ & $0.534\pm0.020$ & SCC & 0.73 & 0.303 & 0.33 \\
A0400 & 44.4110 & 6.0050 & 44.8 & 0.0240 & $2.23_{-0.11}^{+0.09}$ & $0.544_{-0.010}^{+0.010}$ & $126.97_{-4.91}^{+5.06}$ & $0.352\pm0.004$ & NCC & $0.54^{\dagger}$ & 0.371 & 0.48 \\
A0399 & 44.4680 & 13.0440 & 9.3 & 0.0715 & $6.11_{-0.12}^{+0.12}$ & $0.551_{-0.013}^{+0.014}$ & $120.71_{-12.93}^{+13.25}$ & $3.634\pm0.196$ & NCC & 1.20 & 0.051 & 1.36 \\
A0401 & 44.7380 & 13.5790 & 3.7 & 0.0748 & $7.70_{-0.19}^{+0.30}$ & $0.527_{-0.004}^{+0.004}$ & $69.55_{-3.68}^{+3.83}$ & $6.452\pm0.071$ & NCC & 1.30 & 0.346 & 1.42 \\
A3112 & 49.4910 & -44.2370 & 0.5 & 0.0750 & $4.39_{-0.11}^{+0.10}$ & $0.646_{-0.008}^{+0.009}$ & $107.81_{-4.03}^{+4.26}$ & $3.832\pm0.042$ & SCC & 0.99 & 0.246 & 1.42 \\
FORNAX & 54.6490 & -35.3640 & 202.9 & 0.0046 & $1.34_{-0.00}^{+0.00}$ & $0.613_{-0.023}^{+0.025}$ & $101.54_{-4.02}^{+4.12}$ & $0.042\pm0.002$ & SCC & $0.38^{\dagger}$ & 0.061 & 0.09 \\
2A0335 & 54.6660 & 9.9710 & 9.5 & 0.0349 & $3.34_{-0.11}^{+0.09}$ & $0.731_{-0.011}^{+0.012}$ & $141.70_{-4.13}^{+4.23}$ & $2.462\pm0.020$ & SCC & 0.92 & 0.177 & 0.69 \\
IIIZw54 & 55.3150 & 15.3970 & 46.4 & 0.0311 & $2.44_{-0.05}^{+0.04}$ & $0.736_{-0.078}^{+0.120}$ & $166.98_{-34.00}^{+44.75}$ & $0.427\pm0.033$ & WCC & $0.56^{\dagger}$ & 0.104 & 0.62 \\
A3158 & 55.7260 & -53.6300 & 4.0 & 0.0590 & $4.62_{-0.06}^{+0.06}$ & $0.697_{-0.018}^{+0.020}$ & $234.54_{-11.41}^{+11.72}$ & $2.899\pm0.043$ & NCC & 1.12 & 0.207 & 1.14 \\
A0478 & 63.3550 & 10.4660 & 1.3 & 0.0900 & $6.67_{-0.17}^{+0.16}$ & $0.731_{-0.006}^{+0.007}$ & $186.43_{-3.58}^{+3.65}$ & $9.083\pm0.054$ & SCC & 1.31 & 0.240 & 1.68 \\
NGC1550 & 64.9130 & 2.4070 & 37.9 & 0.0123 & $1.34_{-0.00}^{+0.00}$ & $0.615_{-0.038}^{+0.047}$ & $55.14_{-9.43}^{+11.39}$ & $0.155\pm0.008$ & SCC & $0.44^{\dagger}$ & 0.054 & 0.25 \\
EXO0422 & 66.4670 & -8.5620 & 16.4 & 0.0390 & $2.81_{-0.11}^{+0.11}$ & $0.688_{-0.026}^{+0.030}$ & $70.53_{-7.91}^{+8.55}$ & $1.036\pm0.064$ & SCC & $0.69^{\dagger}$ & 0.205 & 0.77 \\
A3266 & 67.8810 & -61.4260 & 86.3 & 0.0594 & $8.52_{-0.32}^{+0.31}$ & $0.933_{-0.018}^{+0.018}$ & $560.18_{-12.26}^{+12.57}$ & $4.483\pm0.031$ & WCC & 1.30 & 0.397 & 1.15 \\
A0496 & 68.4080 & -13.2590 & 8.2 & 0.0328 & $4.50_{-0.05}^{+0.04}$ & $0.651_{-0.010}^{+0.010}$ & $146.49_{-4.67}^{+4.75}$ & $1.972\pm0.014$ & SCC & 0.97 & 0.217 & 0.65 \\
A3376 & 90.5000 & -39.9670 & 52.0 & 0.0455 & $3.58_{-0.09}^{+0.10}$ & $0.675_{-0.026}^{+0.029}$ & $375.21_{-21.94}^{+22.79}$ & $1.118\pm0.016$ & NCC & 0.93 & 0.495 & 0.89 \\
A3391 & 96.5800 & -53.6980 & 30.2 & 0.0531 & $5.30_{-0.32}^{+0.27}$ & $0.762_{-0.034}^{+0.038}$ & $292.76_{-19.82}^{+21.42}$ & $1.379\pm0.026$ & NCC & 0.98 & 0.398 & 1.03 \\
A3395s & 96.6970 & -54.5460 & 11.2 & 0.0498 & $4.47_{-0.23}^{+0.23}$ & $0.699_{-0.086}^{+0.127}$ & $286.63_{-57.77}^{+77.07}$ & $1.096\pm0.042$ & NCC & 1.03 & 0.330 & 0.97 \\
A0576 & 110.3530 & 55.7930 & 103.9 & 0.0381 & $3.83_{-0.09}^{+0.07}$ & $0.716_{-0.072}^{+0.107}$ & $210.87_{-39.80}^{+52.62}$ & $0.962\pm0.065$ & WCC & 0.90 & 0.255 & 0.76 \\
A0754 & 137.3110 & -9.6830 & 82.6 & 0.0528 & $9.99_{-0.38}^{+0.34}$ & $0.767_{-0.015}^{+0.016}$ & $405.13_{-11.60}^{+11.92}$ & $2.052\pm0.033$ & NCC & 1.32 & 0.455 & 1.03 \\
HYDRA-A & 139.5230 & -12.0940 & 2.7 & 0.0538 & $3.27_{-0.08}^{+0.07}$ & $0.719_{-0.007}^{+0.007}$ & $126.94_{-2.63}^{+2.68}$ & $3.049\pm0.018$ & SCC & $0.84^{\dagger}$ & 0.237 & 1.05 \\
A1060 & 159.1680 & -27.5280 & 41.9 & 0.0114 & $3.02_{-0.04}^{+0.04}$ & $0.602_{-0.007}^{+0.007}$ & $98.62_{-1.97}^{+1.92}$ & $0.284\pm0.009$ & WCC & $0.68^{\dagger}$ & 0.062 & 0.23 \\
A1367 & 176.2100 & 19.7030 & 68.4 & 0.0216 & $3.38_{-0.05}^{+0.05}$ & $0.670_{-0.021}^{+0.022}$ & $346.02_{-12.76}^{+13.68}$ & $0.619\pm0.005$ & NCC & 0.83 & 0.089 & 0.44 \\
MKW4 & 181.1020 & 1.8990 & 38.1 & 0.0200 & $2.01_{-0.04}^{+0.04}$ & $0.566_{-0.011}^{+0.011}$ & $79.67_{-3.89}^{+4.05}$ & $0.200\pm0.003$ & SCC & $0.51^{\dagger}$ & 0.275 & 0.41 \\
ZwCl1215 & 184.4200 & 3.6590 & 7.2 & 0.0750 & $5.74_{-0.25}^{+0.28}$ & $0.748_{-0.017}^{+0.018}$ & $248.82_{-10.45}^{+10.78}$ & $2.693\pm0.035$ & NCC & $1.01^{\dagger}$ & 0.268 & 1.42 \\
NGC4636 & 190.7310 & 2.7600 & 272.0 & 0.0037 & $0.90_{-0.02}^{+0.02}$ & $2.043_{-0.682}^{+1.127}$ & $149.01_{-35.75}^{+45.65}$ & $0.012\pm0.001$ & SCC & $0.34^{\dagger}$ & 0.176 & 0.08 \\
A3526 & 192.2300 & -41.3000 & 89.3 & 0.0103 & $3.68_{-0.02}^{+0.02}$ & $0.551_{-0.003}^{+0.003}$ & $110.89_{-1.29}^{+1.30}$ & $0.637\pm0.014$ & SCC & 0.77 & 0.280 & 0.21 \\
A1644 & 194.2900 & -17.3990 & 14.0 & 0.0474 & $4.70_{-0.08}^{+0.08}$ & $0.688_{-0.062}^{+0.084}$ & $296.25_{-54.79}^{+68.92}$ & $1.993\pm0.102$ & SCC & 1.07 & 0.050 & 0.93 \\
A1650 & 194.6700 & -1.7590 & 6.4 & 0.0845 & $5.33_{-0.06}^{+0.05}$ & $0.642_{-0.038}^{+0.048}$ & $171.84_{-26.11}^{+31.73}$ & $3.754\pm0.248$ & WCC & 1.13 & 0.160 & 1.59 \\
A1651 & 194.8390 & -4.1950 & 9.0 & 0.0860 & $5.80_{-0.24}^{+0.24}$ & $0.696_{-0.012}^{+0.012}$ & $180.93_{-6.44}^{+6.75}$ & $4.109\pm0.049$ & WCC & 1.18 & 0.152 & 1.61 \\
COMA & 194.9480 & 27.9300 & 30.6 & 0.0232 & $8.26_{-0.14}^{+0.15}$ & $0.874_{-0.004}^{+0.004}$ & $449.21_{-2.39}^{+2.37}$ & $4.067\pm0.057$ & NCC & 1.35 & 0.295 & 0.47 \\
NGC5044 & 198.8640 & -16.3860 & 38.8 & 0.0090 & $1.22_{-0.04}^{+0.03}$ & $0.664_{-0.013}^{+0.014}$ & $65.97_{-2.13}^{+2.19}$ & $0.099\pm0.000$ & SCC & $0.41^{\dagger}$ & 0.063 & 0.18 \\
A1736 & 201.7500 & -27.1580 & 108.3 & 0.0461 & $2.98_{-0.11}^{+0.10}$ & $0.591_{-0.048}^{+0.062}$ & $337.18_{-56.53}^{+70.00}$ & $1.657\pm0.104$ & NCC & 0.99 & 0.065 & 0.91 \\
A3558 & 202.0050 & -31.5130 & 57.3 & 0.0480 & $4.58_{-0.13}^{+0.11}$ & $0.656_{-0.006}^{+0.007}$ & $265.23_{-4.54}^{+4.67}$ & $3.401\pm0.017$ & WCC & 1.17 & 0.289 & 0.94 \\
A3562 & 203.3960 & -31.6650 & 13.1 & 0.0499 & $4.13_{-0.14}^{+0.18}$ & $0.585_{-0.010}^{+0.011}$ & $192.27_{-7.30}^{+7.73}$ & $1.603\pm0.014$ & WCC & 0.94 & 0.230 & 0.98 \\
A3571 & 206.8670 & -32.8520 & 12.3 & 0.0397 & $6.38_{-0.11}^{+0.11}$ & $0.692_{-0.010}^{+0.010}$ & $199.73_{-5.15}^{+5.33}$ & $4.181\pm0.029$ & WCC & 1.16 & 0.336 & 0.79 \\
A1795 & 207.2190 & 26.5980 & 14.7 & 0.0616 & $5.57_{-0.06}^{+0.06}$ & $0.732_{-0.003}^{+0.003}$ & $166.95_{-1.49}^{+1.52}$ & $5.205\pm0.016$ & SCC & 1.14 & 0.262 & 1.19 \\
A3581 & 211.8790 & -27.0180 & 22.3 & 0.0214 & $1.97_{-0.07}^{+0.07}$ & $0.509_{-0.017}^{+0.018}$ & $25.47_{-4.78}^{+5.36}$ & $0.337\pm0.011$ & SCC & $0.63^{\dagger}$ & 0.210 & 0.43 \\
MKW8 & 220.1690 & 3.4710 & 35.7 & 0.0270 & $2.88_{-0.11}^{+0.11}$ & $0.543_{-0.044}^{+0.061}$ & $100.58_{-26.45}^{+34.75}$ & $0.405\pm0.034$ & NCC & 0.72 & 0.151 & 0.54 \\
A2029 & 227.7310 & 5.7440 & 7.8 & 0.0767 & $7.48_{-0.08}^{+0.08}$ & $0.693_{-0.006}^{+0.006}$ & $160.89_{-3.15}^{+3.15}$ & $8.897\pm0.053$ & SCC & 1.32 & 0.282 & 1.45 \\
A2052 & 229.1880 & 7.0260 & 19.5 & 0.0348 & $3.18_{-0.02}^{+0.02}$ & $0.791_{-0.020}^{+0.022}$ & $163.91_{-6.39}^{+6.83}$ & $1.259\pm0.013$ & SCC & $0.67^{\dagger}$ & 0.207 & 0.69 \\
MKW3S & 230.4660 & 7.7060 & 4.5 & 0.0450 & $3.66_{-0.08}^{+0.08}$ & $0.707_{-0.008}^{+0.009}$ & $102.22_{-2.47}^{+2.57}$ & $1.473\pm0.015$ & SCC & 0.86 & 0.277 & 0.88 \\
A2065 & 230.6080 & 27.7170 & 18.8 & 0.0721 & $4.98_{-0.10}^{+0.18}$ & $1.147_{-0.209}^{+0.467}$ & $510.85_{-103.71}^{+182.02}$ & $2.858\pm0.174$ & WCC & 1.10 & 0.275 & 1.37 \\
A2063 & 230.7790 & 8.6110 & 18.5 & 0.0354 & $3.55_{-0.05}^{+0.05}$ & $0.570_{-0.010}^{+0.011}$ & $60.75_{-4.99}^{+5.21}$ & $1.168\pm0.015$ & WCC & 0.86 & 0.160 & 0.70 \\
A2142 & 239.6000 & 27.2220 & 70.9 & 0.0899 & $7.60_{-0.67}^{+0.88}$ & $0.923_{-0.013}^{+0.013}$ & $556.15_{-10.08}^{+10.25}$ & $10.960\pm0.099$ & WCC & 1.41 & 0.388 & 1.68 \\
A2147 & 240.5640 & 15.9550 & 14.4 & 0.0351 & $3.81_{-0.11}^{+0.10}$ & $0.359_{-0.017}^{+0.021}$ & $85.48_{-19.42}^{+24.73}$ & $1.500\pm0.048$ & NCC & 1.06 & 0.088 & 0.70 \\
A2163 & 243.9410 & -6.1470 & 13.6 & 0.2010 & $14.17_{-0.71}^{+0.71}$ & $0.738_{-0.014}^{+0.015}$ & $368.75_{-13.65}^{+14.01}$ & $17.179\pm0.258$ & NCC & 1.67 & 0.151 & 3.31 \\
A2199 & 247.1580 & 39.5480 & 2.0 & 0.0302 & $4.07_{-0.06}^{+0.06}$ & $0.751_{-0.005}^{+0.005}$ & $182.12_{-1.93}^{+1.96}$ & $2.140\pm0.039$ & SCC & 0.99 & 0.201 & 0.60 \\
A2204 & 248.1960 & 5.5760 & 10.2 & 0.1523 & $8.06_{-0.53}^{+0.63}$ & $0.707_{-0.018}^{+0.020}$ & $173.91_{-10.33}^{+10.89}$ & $13.715\pm0.219$ & SCC & 1.33 & 0.046 & 2.65 \\
A2244 & 255.6740 & 34.0590 & 4.9 & 0.0970 & $5.31_{-0.10}^{+0.09}$ & $0.642_{-0.014}^{+0.015}$ & $126.79_{-7.43}^{+7.91}$ & $4.345\pm0.091$ & WCC & 1.12 & 0.145 & 1.80 \\
A2256 & 255.9750 & 78.6400 & 31.0 & 0.0601 & $6.91_{-0.55}^{+0.57}$ & $1.041_{-0.013}^{+0.014}$ & $547.10_{-7.35}^{+7.58}$ & $4.793\pm0.067$ & NCC & 1.27 & 0.341 & 1.16 \\
A2255 & 258.2080 & 64.0650 & 26.5 & 0.0800 & $5.33_{-0.18}^{+0.17}$ & $0.851_{-0.027}^{+0.029}$ & $502.79_{-20.19}^{+21.65}$ & $2.829\pm0.034$ & NCC & 1.21 & 0.209 & 1.51 \\
A3667 & 303.1450 & -56.8430 & 17.3 & 0.0560 & $5.84_{-0.04}^{+0.04}$ & $0.567_{-0.006}^{+0.007}$ & $260.96_{-6.82}^{+7.05}$ & $4.949\pm0.035$ & WCC & 1.33 & 0.341 & 1.09 \\
S1101 & 348.4940 & -42.7270 & 1.3 & 0.0580 & $2.50_{-0.11}^{+0.11}$ & $0.783_{-0.011}^{+0.011}$ & $115.72_{-3.15}^{+3.27}$ & $1.850\pm0.017$ & SCC & 0.79 & 0.178 & 1.12 \\
A2589 & 350.9880 & 16.7720 & 10.7 & 0.0416 & $3.65_{-0.04}^{+0.04}$ & $0.671_{-0.016}^{+0.017}$ & $148.38_{-6.86}^{+7.35}$ & $0.989\pm0.013$ & WCC & 0.84 & 0.285 & 0.82 \\
A2597 & 351.3320 & -12.1250 & 1.9 & 0.0852 & $3.79_{-0.06}^{+0.06}$ & $0.702_{-0.013}^{+0.014}$ & $106.77_{-5.02}^{+5.42}$ & $3.535\pm0.042$ & SCC & 0.93 & 0.192 & 1.60 \\
A2634 & 354.6090 & 27.0130 & 61.9 & 0.0312 & $3.04_{-0.10}^{+0.10}$ & $0.680_{-0.027}^{+0.030}$ & $297.03_{-16.72}^{+17.64}$ & $0.518\pm0.008$ & WCC & 0.80 & 0.554 & 0.62 \\
A2657 & 356.2290 & 9.1940 & 16.4 & 0.0404 & $3.33_{-0.10}^{+0.11}$ & $0.555_{-0.006}^{+0.006}$ & $96.03_{-3.00}^{+3.05}$ & $0.911\pm0.008$ & WCC & 0.79 & 0.125 & 0.80 \\
A4038 & 356.9320 & -28.1450 & 13.3 & 0.0283 & $3.00_{-0.04}^{+0.03}$ & $0.620_{-0.013}^{+0.014}$ & $98.38_{-4.88}^{+5.08}$ & $1.005\pm0.013$ & WCC & 0.80 & 0.313 & 0.57 \\
A4059 & 359.2520 & -34.7580 & 7.1 & 0.0460 & $3.94_{-0.03}^{+0.03}$ & $0.705_{-0.018}^{+0.020}$ & $168.53_{-8.62}^{+9.33}$ & $1.477\pm0.019$ & SCC & 0.94 & 0.255 & 0.90 \\
\hline
\end{longtable}
\tablefoot{$^{\dagger}$}
\end{landscape}
\twocolumn

%
\includepdf[pages=-]{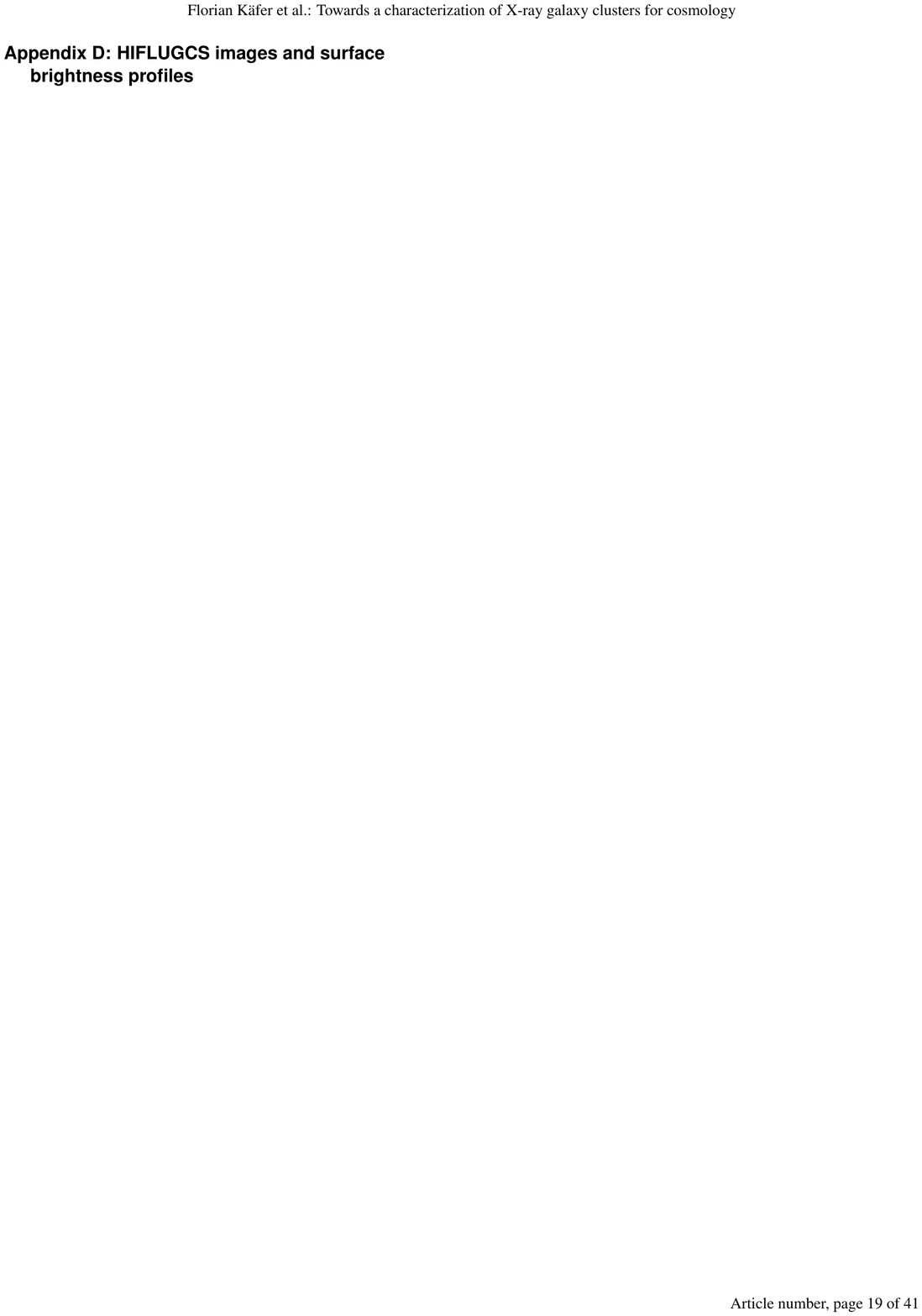}

\end{appendix}

\end{document}